\documentclass[12pt]{iopart}

\usepackage{bbold}
\usepackage{braket}
\expandafter\let\csname equation*\endcsname\relax
\expandafter\let\csname endequation*\endcsname\relax
\expandafter\let\csname Tr\endcsname\relax
\expandafter\let\csname table*\endcsname\relax
\expandafter\let\csname tabular\endcsname\relax
\expandafter\let\csname ruledtabular\endcsname\relax
\expandafter\let\csname endruledtabular\endcsname\relax
\expandafter\let\csname endtabular\endcsname\relax
\expandafter\let\csname endtable*\endcsname\relax

\usepackage{amsmath}
\usepackage{hyperref}
\usepackage{ulem}
\usepackage{booktabs}
\usepackage[caption=false]{subfig}
\usepackage{braket}
\usepackage{dcolumn}
\usepackage[dvipsnames]{xcolor}

\usepackage{graphicx}

\usepackage{floatrow}
\floatstyle{plaintop}
\restylefloat{table}
\usepackage{subfig}

\DeclareMathOperator{\Tr}{Tr}

\begin{document}

\title[Two-mode Schr{\"o}dinger-cat states with nonlinear optomechanics]{Two-mode Schr{\"o}dinger-cat states with nonlinear optomechanics: generation and verification of non-Gaussian mechanical entanglement}

\author{Lydia A. Kanari-Naish$^1$, Jack Clarke$^1$, 
Sofia Qvarfort$^{1,2,3,4}$ and Michael R. Vanner$^1$}

\address{$^1$ QOLS, Blackett Laboratory, Imperial College London, London SW7 2BW, United Kingdom}
\address{$^2$ Department of Physics and Astronomy, University College London, London WC1E 6BT, United Kingdom}
\address{$^3$ Nordita, KTH Royal Institute of Technology and Stockholm University, Hannes Alfv\'{e}ns v\"{a}g 12, SE-106 91 Stockholm, Sweden}
\address{$^4$ Department of Physics, Stockholm University, AlbaNova University Center, SE-106 91 Stockholm, Sweden}
\ead{l.kanari-naish18@imperial.ac.uk}
\ead{m.vanner@imperial.ac.uk (www.qmeas.net) }

\begin{abstract}
Cavity quantum optomechanics has emerged as a new platform for quantum science and technology with applications ranging from quantum-information processing to tests of the foundations of physics. Of crucial importance for optomechanics is the generation and verification of non-Gaussian states of motion and a key outstanding challenge is the observation of a canonical two-mode Schr{\"o}dinger-cat state in the displacement of two mechanical oscillators.
In this work, we introduce a pulsed approach that utilizes the nonlinearity of the radiation-pressure interaction combined with photon-counting measurements to generate this entangled non-Gaussian mechanical state, and, importantly, describe a protocol using subsequent pulsed interactions to verify the non-Gaussian entanglement generated.
Our pulsed verification protocol allows quadrature moments of the two mechanical oscillators to be measured up to any finite order providing a toolset for experimental characterisation of bipartite mechanical quantum states and allowing a broad range of inseparability criteria to be evaluated.
Key experimental factors, such as optical loss and open-system dynamics, are carefully analyzed and we show that the scheme is feasible with only minor improvements to current experiments that operate outside the resolved-sideband regime. Our scheme provides a new avenue for quantum experiments with entangled mechanical oscillators and offers significant potential for further research and development that utilizes such non-Gaussian states for quantum-information and sensing applications, and for studying the quantum-to-classical transition.
\end{abstract}

\vspace{2pc}

\maketitle

\section{\label{sec:Introduction}Introduction}
Quantum mechanics allows for two or more objects to exhibit correlations that are stronger than is permitted by classical physics. Such correlations---or quantum entangled states---are one of the most counter-intuitive and powerful aspects of quantum mechanics, and offer exciting routes to develop quantum technologies and to explore fundamental physics. Indeed, quantum entanglement provides significant potential to surpass limitations set by classical physics for widespread applications, including quantum communications~\cite{Ekert1991}, computing~\cite{Gottesman1999,Menicucci2006}, networking~\cite{Briegel1998,Wehner2018}, and sensing~\cite{Giovannetti2011, Joo2011}. Additionally, entanglement is central to many studies of fundamental physics with a prominent example being tests of Bell nonlocality~\cite{Hensen2015, Shalm2015, Giustina2015}.

With cavity quantum optomechanics now providing a rapidly progressing new platform for quantum science~\cite{Aspelmeyer2014}, quantum entanglement in macroscopic mechanical oscillators is currently emerging as an active avenue of study. In particular, the radiation-pressure optomechanical interaction can be utilized to generate a rich variety of different entangled states. 
Notably, this interaction, when linearized, gives rise to Gaussian entanglement between optical and mechanical modes, which has been studied both theoretically and experimentally. Theoretically, entanglement between both field and mechanics, and also between two mechanical elements has been explored in this regime~\cite{Mancini2002, Pirandola2006, Wang2013, Woolley2014, Vostrosablin2016, Clarke2020, Brunelli2020, Brandao2020, Gut2020, Neveu2021}. While experimentally, two-mode squeezed states have been prepared between a microwave field and a mechanical oscillator~\cite{Palomaki2013}, and excitingly, continuous-variable entanglement between two mechanical oscillators has also been generated~\cite{Ockeloen-Korppi2018, Lepinay2021, Kotler2021}.
In addition to such Gaussian states, the linearized regime combined with photon-counting enables the preparation of single- and two-mode non-Gaussian mechanical states~\cite{Borkje2011, Vanner2013, Flayac2014, Milburn2016}. Recent experimental progress using this combination has enabled single-phonon addition and subtraction to single-mode mechanical systems~\cite{Lee2012, Riedinger2016, Fisher2017,Velez2019,Enzian2021, Patel2021, Enzian2021nG}, and, similarly, to generate entangled mechanical states that share a single quanta~\cite{Lee2011, Riedinger2018}. Furthermore, photon-counting protocols have been proposed to create macroscopic superposition states which can help to overcome the challenge of single-photon weak coupling~\cite{Pepper2012, Ringbauer2018}.
Beyond the linearized regime, continued experimental progress has enabled early signatures of the intrinsically cubic radiation-pressure interaction to be observed~\cite{Brawley2016, Leijssen2017}, and there has also been increasing theoretical interest in this nonlinear regime~\cite{Mancini1997, Bose1997, Rabl2011, Nunnenkamp2011, Qvarfort2021}. 
Theoretical progress in this direction has examined entangling operations and the properties of the mechanical states generated~\cite{Vacanti2008, Akram2013, Xiong2019}, and the nonlinear regime remains as a rich avenue of study for mechanical quantum states. 
In particular, an experimental recipe for mechanical continuous-variable non-Gaussian entanglement verification remains outstanding. 

In this work, we propose how to generate two-mode mechanical Schr{\"o}dinger-cat states encoded in the displacement of two mechanical oscillators, and importantly introduce an operational technique to verify the non-Gaussian entanglement with a scheme that enables any bipartite mechanical moment to be measured up to any finite order. Our protocol comprises a state preparation stage followed by a verification stage, which both utilize pulsed optomechanical interactions~\cite{Vanner2011} in the unresolved sideband regime to provide a quantum-non-demolition-type interaction with the mechanical position quadrature.
For entanglement generation, we take advantage of the nonlinear radiation-pressure interaction together with photon-counting measurements with an interferometric set-up to herald a two-mode mechanical Schr{\"o}dinger-cat state. For the second entanglement verification stage, we propose a method to determine arbitrary mechanical moments utilizing subsequent pulsed measurements with an iterative verification protocol whereby lower-order moments are used to unlock higher-order moments. We then use the moments obtained to evaluate inseparability criteria which identify the presence of non-Gaussian mechanical-mechanical entanglement~\cite{Shchukin2005}.

Our scheme is applicable to current experimental approaches and non-Gaussian entanglement can be created and verified with only minor improvements to current experiments operating outside the resolved sideband regime. 
To establish the feasibility of this scheme, we carefully model the key experimental factors including those arising from the system's interaction with the environment which occur during the verification stage; and optical losses, detection inefficiency, and dark counts in the heralding stage. These specific entangled states generated by our scheme have a wide range of applications with prominent examples including quantum metrology~\cite{Joo2011}, quantum teleportation~\cite{vanEnk2001}, quantum networks~\cite{Fiaschi2021}, and fault-tolerant quantum computation~\cite{Wang2016}. More broadly, these entangled states can also be applied for empirical studies of quantum macroscopicity~\cite{LeeCW2011,Yadin2016}, sensing, and the quantum-to-classical transition~\cite{Bose1999, Marshall2003, Kleckner2008, Bassi2013}.
Furthermore, the verification protocol we introduce can be used to assess entanglement criteria and perform moment-based state characterisation for any bipartite mechanical quantum state of motion.

\section{\label{sec:Entanglement_Protocol} Entanglement Protocol}

\begin{figure*}
\centering
  \sidesubfloat[]%{\includegraphics[width=0.35\linewidth, trim = 10mm 0mm -10mm 0mm]{parallel_setup_diagram.pdf}
  {\includegraphics[height=2.7cm]{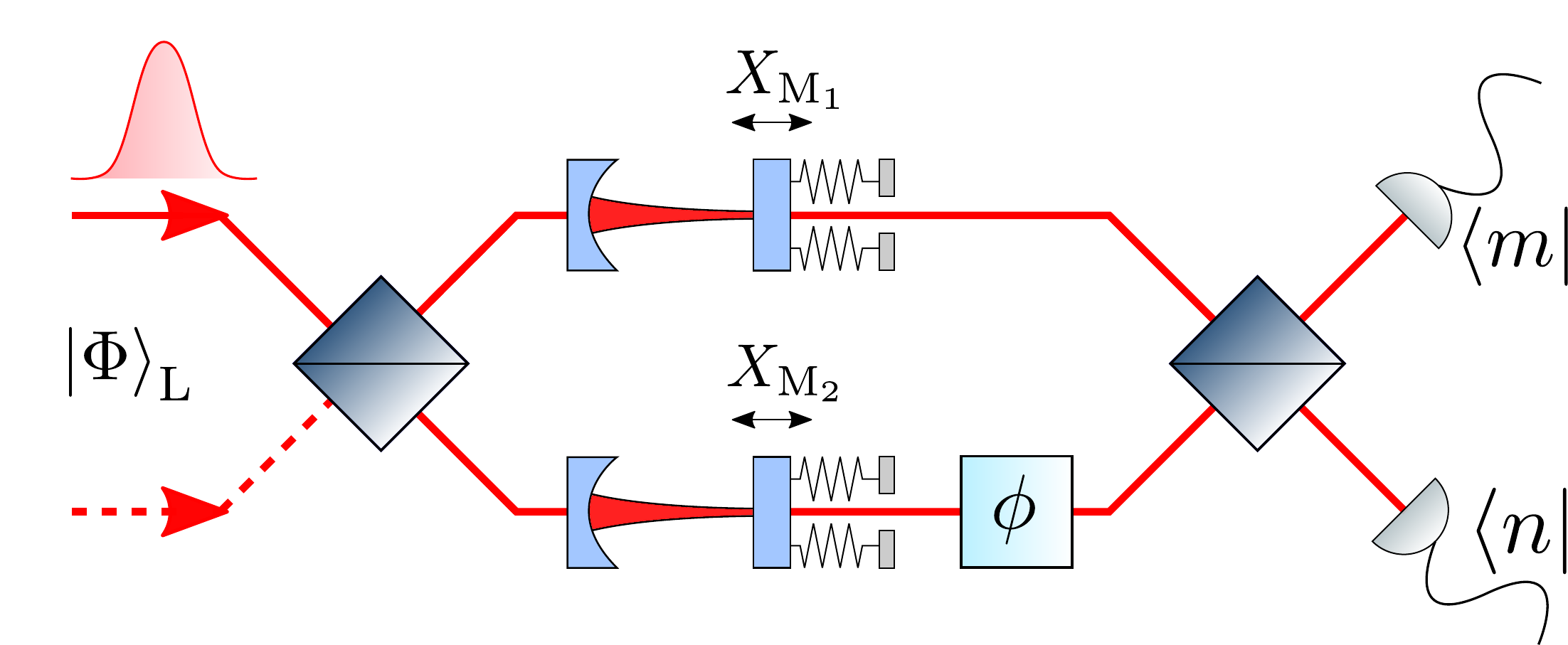}
  \label{fig:parallel}}
  %$\qquad$
  \sidesubfloat[]%{\includegraphics[width=0.4\linewidth, trim = 10mm 0mm -10mm 0mm]{series_setup_diagram.pdf}
  {\includegraphics[height=2.7cm]{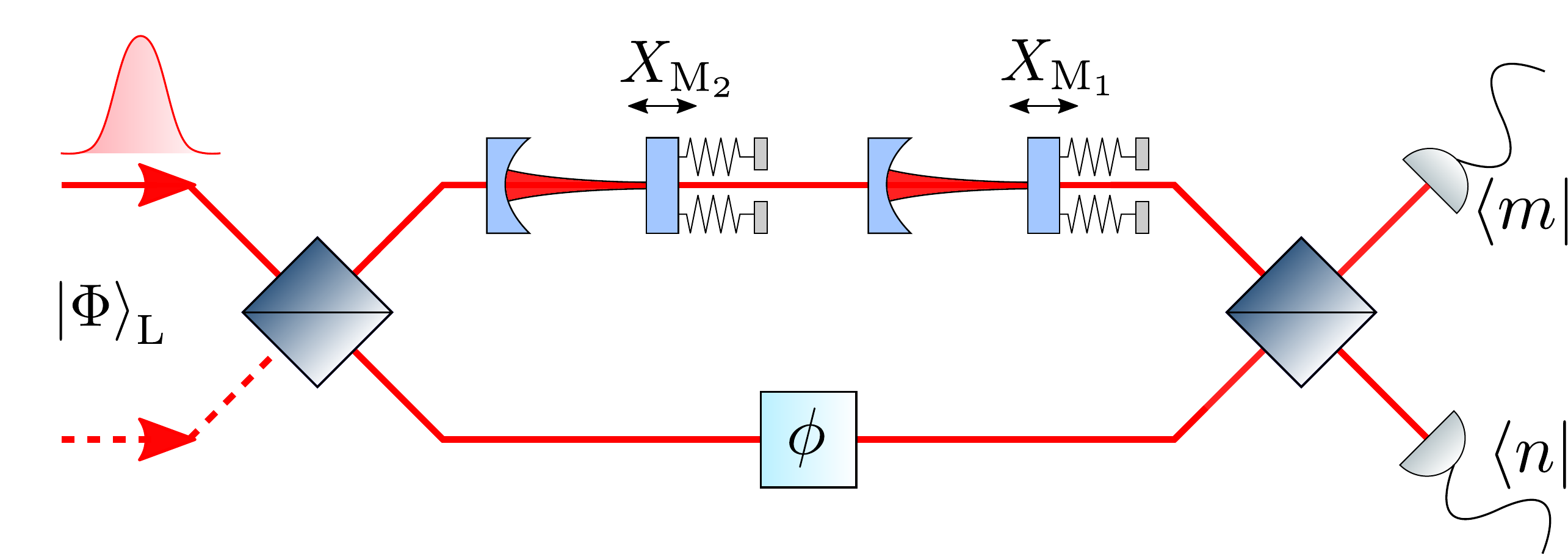}
  \label{fig:series}}%
  \caption{Proposed experimental schematics to prepare mechanical oscillators in a two-mode Schr{\"o}dinger-cat state. Both schemes involve injecting a weak coherent state $\ket{\Phi}_{\mathrm{L}}=\ket{\alpha}_{\mathrm{L}_1}\ket{0}_{\mathrm{L}_2}$, or a single photon $\ket{\Phi}_{\mathrm{L}}=\ket{1}_{\mathrm{L}_1}\ket{0}_{\mathrm{L}_2}$ into a Mach-Zehnder interferometer containing two optomechanical cavities. Following a nonlinear interaction, photon-counting is then performed at the two outputs of the interferometer, with counts $\{m,n\}$, that heralds the creation of the mechanical state $\rho^{\{m,n\}}_{\mathrm{out}}$. The phase $\phi$ influences the properties of the state. (a) The `parallel' set-up described by the measurement operator in Eq. \eqref{eq:Upsilon1}. (b) The `series' set-up described by the measurement operator in Eq. \eqref{eq:Upsilon2}. The phase $\phi$ in both configurations is used to control the properties of the resulting non-Gaussian entangled mechanical state.}
  \label{fig:entanglement_protocols}
\end{figure*}

We first outline our scheme to generate non-Gaussian entanglement between two mechanical oscillators before exploring each stage in more mathematical detail. This entanglement scheme can be applied to optomechanical systems that operate outside the resolved sideband regime $\kappa\gg\omega{_\textsc{m}}$ (where $\kappa$ is the cavity amplitude decay rate and $\omega_\textsc{m}$ is the mechanical angular frequency). We consider two distinct configurations involving a Mach-Zehnder interferometer set-up: (i) the `parallel' case where light in the two paths interact with separate mechanical oscillators (see Fig. \ref{fig:parallel}), and (ii) the `series' case where only light in the first path interacts with two mechanical oscillators sequentially (see Fig. \ref{fig:series}). 
In both cases, a pulse of light is first injected into one input of the interferometer. The light in the lower interferometer-path is subjected to a phase shift $\phi$, which is an experimental handle that allows us to directly influence the the properties of the resulting mechanical quantum states. Following a pulsed nonlinear optomechanical interaction \cite{Pikovski2012,Wang2017,Clarke2018}, the two optical modes recombine at another 50:50 beam splitter and are subsequently measured by photon-counting detectors. We denote the event where $m$ and $n$ photons are detected in the two outputs as $\{m,n\}$ thus heralding the creation of the entangled mechanical state $\rho^{\{m,n\}}_{\mathrm{out}}$.

\floatsetup[figure]{style=plain,subcapbesideposition=top}

We consider separately two initial optical states to implement our scheme. Firstly, a weak coherent state, i.e. $\ket{\Phi}_{\mathrm{L}}=\ket{\alpha}_{\mathrm{L}_1}\ket{0}_{\mathrm{L}_2}$, where subscripts 1 and 2 denote upper and lower light modes in the interferometer. And, secondly, a single-photon input, i.e. $\ket{\Phi}_{\mathrm{L}}=\ket{1}_{\mathrm{L}_1}\ket{0}_{\mathrm{L}_2}$ which offers advantages for the quantum measurement process, however with the increased experimental complexity of requiring a single-photon source. At the first beam splitter the light interacts with the vacuum mode and undergoes the following transformations: $U_{\mathrm{12}}^\dag a_1 U_{\mathrm{12}}\rightarrow (a_1+a_2)/\sqrt{2}$ and $U_{\mathrm{12}}^\dag a_2 U_{\mathrm{12}}\rightarrow (a_1-a_2)/\sqrt{2}$, where $a_{1}$ and $a_2$ are the annihilation operators associated with optical modes 1 and 2. The subsequent interaction between the optical pulse and the mechanical oscillator is captured by the following Hamiltonian:

\begin{equation}
    \label{eq:hamiltonian}
    H_\mathrm{int}=-\hbar g_0 a_i^\dag a_i (b_j+b_j^\dag)~,
\end{equation}
where $g_0$ is the optomechanical coupling rate, and $b_j$ is the mechanical annihilation operator for the $j^{\mathrm{th}}$ oscillator. We do not require strong coupling and furthermore assume the same $g_0$ for the two optomechanical cavities for brevity and clarity of presentation, but that can be readily generalized if needed. In this pulsed regime, the pulse duration is much shorter than the mechanical period allowing us to model the optomechanical interaction by the unitary $\mathrm{e}^{\mathrm{i}\mu a_i^\dag a_i X_{\mathrm{M}_j}}$; where $\mu \propto g_0/\kappa$ is the dimensionless coupling strength (and depends on the shape of the pulse), and $X_{\mathrm{M}_j}=(b_j+b_j^\dag)/\sqrt{2}$ is the position-quadrature operator of the $j^\mathrm{th}$ mechanical system. We remind the reader that the pulsed regime assumes $\kappa\gg\omega_{\mathrm{M}}$, however, the interplay between the unitary $\mathrm{e}^{\mathrm{i}\mu a^\dag_i a_i X_{\mathrm{M}_j}}$ and a finite value of the ratio $\omega_{\mathrm{M}}/\kappa$ is analysed in \ref{app:sideband_res}, where we demonstrate that the entanglement generation protocol is robust with respect to this ratio.

For a pulse duration $\tau$ which satisfies $\omega_{\mathrm{M}}\ll \tau^{-1}\ll \kappa$, it can be shown that $\mu=2\sqrt{2}g_0/\kappa$ as the cavity mode can be adiabatically eliminated~\cite{Pikovski2012}. We also assume that the input pulse shape is Gaussian with temporal width $\tau\gg1/\kappa$, as this allows one to adiabatically eliminate the cavity field even at early times in the cavity's dynamics~\cite{khosla2013quantum,Bennett2016}. 
This value of $\mu=2\sqrt{2}g_0/\kappa$ is the result of the coherent sum over all possible trajectories the photons may take as they enter and then leave the cavity. Certain trajectories where the photons reflect only a small number of times from the mechanical mode, and trajectories where the photons remain in the cavity for a very long time compared to $1/\kappa$, lead to smaller or larger optomechanical couplings, respectively. However, in the adiabatic regime these weakly-interacting and strongly-interacting photon trajectories occur with a negligible probability amplitude and so the optomechanical coupling strength is well approximated by $\mu=2\sqrt{2}g_0/\kappa$.
Furthermore, the input-output theory~\cite{gardiner1985input} used to derive  $\mathrm{e}^{\mathrm{i}\mu a^\dag_i a_i X_{\mathrm{M}_j}}$ in Ref.~\cite{Pikovski2012} demonstrates that in this adiabatic regime the output pulse from the optomechanical cavity experiences minimal distortion. This has the advantage of more easily ensuring spatio-temporal mode matching at the second beam splitter if the two optomechanical cavities are not totally identical, e.g. slightly different values of $\kappa$ between the two cavities.

An $\{m,n\}$ click heralds the mechanical state $\rho^{\{m,n\}}_{\mathrm{out}}=\Upsilon_{mn}\rho_{\mathrm{in}}\Upsilon^\dag_{mn}/{P}_{mn}$, where $\rho_{\mathrm{in}}$ is the initial mechanical density operator, and ${P}_{mn}$ is the heralding probability of an $\{m,n\}$ click event. The measurement operator $\Upsilon_{mn}$ captures the propagation of the light through the interferometer, its interaction with the mechanical systems, and the final measurement of $\{m,n\}$ photons. 
For the parallel configuration, this is given by $\Upsilon_{mn}=\bra{m}\bra{n} U_{\mathrm{12}}^\dag \mathrm{e}^{\mathrm{i}\mu a_1^\dag a_1 X_{\mathrm{M}_1}}\mathrm{e}^{\mathrm{i}\mu a_2^\dag a_2 X_{\mathrm{M}_2}}\mathrm{e}^{\mathrm{i}\phi a_2^\dag a_2}U_{\mathrm{12}}\ket{\Phi}_{\mathrm{L}}$, where $\ket{m}$ and $\ket{n}$ are Fock states. The measurement operator for the series configuration can be deduced in a similar manner. For $\ket{\Phi}_{\mathrm{L}}=\ket{\alpha}_{\mathrm{L}_1}\ket{0}_{\mathrm{L}_2}$, the measurement operators are:

\begin{subequations}
\begin{align}
    \Upsilon_{mn}&=
      \mathcal{N}_{mn}[\mathrm{e}^{\mathrm{i}\mu X_{\mathrm{M}_1}}+\mathrm{e}^{\mathrm{i}\mu X_{\mathrm{M}_2}+\mathrm{i}\phi}]^m[\mathrm{e}^{\mathrm{i}\mu X_{\mathrm{M}_1}}-\mathrm{e}^{\mathrm{i}\mu X_{\mathrm{M}_2}+\mathrm{i}\phi}]^n,\label{eq:Upsilon1}\\
     \Upsilon_{mn}&=\mathcal{N}_{mn}[\mathrm{e}^{\mathrm{i}\mu (X_{\mathrm{M}_1}+X_{\mathrm{M}_2})}+\mathrm{e}^{\mathrm{i}\phi}]^m[\mathrm{e}^{\mathrm{i}\mu (X_{\mathrm{M}_1}+X_{\mathrm{M}_2})}-\mathrm{e}^{\mathrm{i}\phi}]^n,\label{eq:Upsilon2}
\end{align}
\end{subequations} 
for the parallel and series set-ups, respectively. For a single-photon input state $\ket{\Phi}_{\mathrm{L}}=\ket{1}_{\mathrm{L}_1}\ket{0}_{\mathrm{L}_2}$, the measurement operators are:
\begin{subequations}
\begin{align}
\begin{split}
    \Upsilon_{mn}&=
      \mathcal{N}_{mn}[(\mathrm{e}^{\mathrm{i}\mu X_{\mathrm{M}_1}}+\mathrm{e}^{\mathrm{i}\mu X_{\mathrm{M}_2}+\mathrm{i}\phi})\delta_{m,1}\delta_{n,0}+(\mathrm{e}^{\mathrm{i}\mu X_{\mathrm{M}_1}}-\mathrm{e}^{\mathrm{i}\mu X_{\mathrm{M}_2}+\mathrm{i}\phi})\delta_{m,0}\delta_{n,1}],\label{eq:Upsilon11}
      \end{split}\\
      \begin{split}
     \Upsilon_{mn}&=\mathcal{N}_{mn}[(\mathrm{e}^{\mathrm{i}\mu (X_{\mathrm{M}_1}+X_{\mathrm{M}_2})}+\mathrm{e}^{\mathrm{i}\phi})\delta_{m,1}\delta_{n,0}+(\mathrm{e}^{\mathrm{i}\mu (X_{\mathrm{M}_1}+X_{\mathrm{M}_2})}-\mathrm{e}^{\mathrm{i}\phi})\delta_{m,0}\delta_{n,1}],
     \end{split}
     \label{eq:Upsilon22}
\end{align}
\end{subequations} 
for the parallel and series cases, respectively. The prefactor $\mathcal{N}_{mn}$ depends on the injected light state $\ket{\Phi}_{\mathrm{L}}$:
\begin{align}
\mathcal{N}_{mn}=
\begin{cases}
      \mathrm{e}^{-\frac{|\alpha|^2}{2}}\left(\frac{\alpha}{2}\right)^{m+n}\frac{1}{\sqrt{n!m!}} & \text{for} \ket{\Phi}_{\mathrm{L}}=\ket{\alpha}_{\mathrm{L}_1}\ket{0}_{\mathrm{L}_2},\\
      1/2 & \text{for} \ket{\Phi}_{\mathrm{L}}=\ket{1}_{\mathrm{L}_1}\ket{0}_{\mathrm{L}_2}.
    \end{cases} 
\end{align}\label{eq:N}
Two-mode mechanical Schr{\"o}dinger-cat states of interest are heralded by the click events $\{1,0\}$ or $\{0,1\}$, although we note that with high efficiency and photon-resolving detectors, $\{m,n\}$ photon click events can be exploited to herald a range of other mechanical states. 
In fact, the operators $\Upsilon_{10}$ and $\Upsilon_{01}$ can be made identical with a $\pi$ phase shift in the interferometer. As $\phi$ is an experimental control, we only need to consider one of these operators to probe the entanglement structure of the state.
Therefore, without loss of generality, subsequent analysis will focus on the click event $\{1,0\}$, which heralds the mechanical state $\rho^{\{1,0\}}_{\mathrm{out}}$. 

As a motivating example, let us consider the case where the mechanical oscillators are initially in their ground states $\ket{0}_{\mathrm{M}_1}\ket{0}_{\mathrm{M}_2}$, and an optical coherent state is used $\ket{\Phi}_{\mathrm{L}}=\ket{\alpha}_{\mathrm{L}_1}\ket{0}_{\mathrm{L}_2}$. After the input pulse travels through the first beam splitter the resulting product state is $\ket{\alpha/\sqrt{2}}_{\mathrm{L}_1}\ket{\mathrm{e}^{\mathrm{i}\phi}\alpha/\sqrt{2}}_{\mathrm{L}_2}$, where the lower path has also experienced a phase shift. In the parallel set-up shown in Fig. \ref{fig:parallel}, the upper path L1 leads to an interaction with oscillator M1 while the lower path L2 interacts with oscillator M2. The total light-mechanics state is now proportional to $D_{\mathrm{L}_1}(\mathrm{e}^{\mathrm{i}\mu X_{\mathrm{M}_1}}\alpha/\sqrt{2})D_{\mathrm{L}_2}(\mathrm{e}^{\mathrm{i}\mu X_{\mathrm{M}_2}+\mathrm{i}\phi}\alpha/\sqrt{2})\ket{0}_{\mathrm{L}_1}\ket{0}_{\mathrm{L}_2}\ket{0}_{\mathrm{M}_1}\ket{0}_{\mathrm{M}_2}$, where $D_{\mathrm{L}_j}$ is the quantum optics displacement operator acting on the $j^{\mathrm{th}}$ optical mode.
At the second beam splitter the light beams are recombined yielding the state
$D_{\mathrm{L}_1}((\mathrm{e}^{\mathrm{i}\mu X_{\mathrm{M}_1}}+\mathrm{e}^{\mathrm{i}\mu X_{\mathrm{M}_2}+\mathrm{i}\phi})\alpha/2)D_{\mathrm{L}_2}((\mathrm{e}^{\mathrm{i}\mu X_{\mathrm{M}_1}}-\mathrm{e}^{\mathrm{i}\mu X_{\mathrm{M}_2}+\mathrm{i}\phi})\alpha/2)\ket{0}_{\mathrm{L}_1}\ket{0}_{\mathrm{L}_2}\ket{0}_{\mathrm{M}_1}\ket{0}_{\mathrm{M}_2}$. We cannot infer the mechanical state yet until photon measurement is performed.
The detection of a single photon in one detector but no photon in the other detector within the coincidence window dictated by the pulse duration gives the desired $\{1,0\}$ click, while we discard runs which have different click events. A $\{1,0\}$ click projects the mechanical state into the form $\mathrm{e}^{\mathrm{i}\mu X_{\mathrm{M}_1}}\ket{0}_{\mathrm{M}_1}\ket{0}_{\mathrm{M}_2}+\mathrm{e}^{\mathrm{i}\mu X_{\mathrm{M}_2}+\mathrm{i}\phi}\ket{0}_{\mathrm{M}_1}\ket{0}_{\mathrm{M}_2}$. Since $\mu$ is a real number, the operators $\mathrm{e}^{\mathrm{i}\mu X_{\mathrm{M}_j}}$ can be expressed as $\exp[{(\mathrm{i}\mu/\sqrt{2}) b_j^\dag-(\mathrm{i}\mu/\sqrt{2})^*b_j}]$ which is the usual displacement operator $D_{\mathrm{M}_j}(\mathrm{i}\mu/\sqrt{2})$. These displacement operators act on the mechanical ground state of the $j^{\mathrm{th}}$ oscillator in the usual manner: $D_{\mathrm{M}_j}(\mathrm{i}\mu/\sqrt{2})\ket{0}_{\mathrm{M}_j}=\ket{\mathrm{i}\mu/\sqrt{2}}_{\mathrm{M}_j}$. As the amplitude of this displacement is purely imaginary, this physically corresponds to a momentum kick of magnitude $\mu$. Therefore, the heralded mechanical states in the parallel and series set-ups, respectively, are given by the two-mode Schr\"{o}dinger-cat states:
\begin{subequations}
\begin{align}
    \ket{\Psi}_{10}\propto
    \left(\ket{\frac{\mathrm{i}\mu}{\sqrt{2}}}\ket{0}+\mathrm{e}^{\mathrm{i}\phi}\ket{0}\ket{\frac{\mathrm{i}\mu}{\sqrt{2}}}\right),\label{eq:pure_state_1}\\
    \ket{\Psi}_{10}\propto\left(\ket{\frac{\mathrm{i}\mu}{\sqrt{2}}}\ket{\frac{\mathrm{i}\mu}{\sqrt{2}}}+\mathrm{e}^{\mathrm{i}\phi}\ket{0}\ket{0}\right)\label{eq:pure_state_2},
\end{align}
\end{subequations}
and are independent of which input optical pulse $\ket{\Phi}_{\mathrm{L}}$ is used (for convenience we have dropped the labels M1 and M2 when describing the bipartite mechanical state). In the parallel example we have been considering, one oscillator receives a single-photon momentum kick while the other remains in its ground state but whichway information has been erased.

Two-mode Schr{\"o}dinger-cat states are usually defined as being proportional to $\ket{\alpha}\ket{\beta}\pm\ket{-\alpha}\ket{-\beta}$ where $\alpha, \beta \in \mathbb{C}$. Acting upon such a state with the local displacement operations $D_{\mathrm{M}_1}(\alpha)D_{\mathrm{M}_2}(-\beta)$ and setting $2\alpha = -2\beta = \mathrm{i}\mu/\sqrt{2}$ gives rise to Eq. \ref{eq:pure_state_1}, while $D_{\mathrm{M}_1}(\alpha)D_{\mathrm{M}_2}(\beta)$ with $2\alpha = 2\beta =\mathrm{i}\mu/\sqrt{2}$ gives Eq. \ref{eq:pure_state_2}. Again, we denote the displacement operator acting on the $j^{\mathrm{th}}$ oscillator as $D_{\mathrm{M}_j}=\mathrm{e}^{\alpha b_j^\dag -\alpha^* b_j}$. By examining these two equations we can see that
since $\braket{0|\mathrm{i}\mu/\sqrt{2}}=\mathrm{e}^{-\mu^2/4}$, for large $\mu$ the two components become increasingly orthogonal, and approach maximally-entangled Bell states in the coherent basis. We also note that for $\mu\ll 1$ and with $\phi=\pi$, we can expand Eqs. \eqref{eq:pure_state_1} and \eqref{eq:pure_state_2} in the Fock basis. This also results in the Bell states $\ket{\Psi^{\pm}}=(\ket{0}\ket{1}\pm\ket{1}\ket{0})/\sqrt{2}$ where $\ket{\Psi^{-}}$ corresponds to the parallel state in Eq. \eqref{eq:pure_state_1} and $\ket{\Psi^{+}}$ the series state in \eqref{eq:pure_state_2}. Indeed, a calculation of the von Neumann entanglement entropy~\cite{Nielsen_Chuang_2010} confirms that for a given value of $\mu$ the entanglement is optimized when $\phi=\pi$.

We now turn our attention to an initial thermal state $\rho_\mathrm{in}=\rho_{\bar{n}_1}\otimes\rho_{\bar{n}_2}$, where $\rho_{\bar{n}_j}$ denotes the $j^{\mathrm{th}}$ oscillator being in a thermal state with thermal occupation number $\bar{n}_j$. The resulting state $\rho_{\mathrm{out}}^{\{1,0\}}$ approaches the desired two-mode Schr{\"o}dinger-cat state as $\bar{n}_{1}$ and $\bar{n}_{2}$ are reduced and the probability of generating $\rho^{\{1,0\}}_{\mathrm{out}}$ is:

\begin{equation}
\label{eq:heralding_prob}
{P}_{10}=\mathcal{N}_{10}[1+\mathrm{e}^{-\mu^2(1+\bar{n}_1+\bar{n}_2)/2}\cos(\phi)].
\end{equation} 
With a coherent state input pulse, ${P}_{10}$ is maximised when $|\alpha|=1$. We note that the mechanical states produced by the parallel (Fig. \ref{fig:parallel}) and series schemes (Fig. \ref{fig:series})  are related by a local unitary operation. Acting on the bi-partite mechanical state produced in the parallel set-up with the unitary $D_{\mathrm{M}_2}(\mathrm{i}\mu/\sqrt{2})R_{\mathrm{M}_2}(\pi)$ will map the state onto the density matrix produced via the series set-up, where $R_{\mathrm{M}_2}(\pi)=\mathrm{e}^{-\mathrm{i}\pi b_2^\dag b_2}$. Therefore, the von Neumann entropies of entanglement of the states produced via the parallel and series configurations are identical \cite{Guifre2000}. Nevertheless, implementing the parallel set-up may be more experimentally convenient as it more easily ensures temporal-mode matching at the beam splitter. Therefore, subsequent analysis will consider the entangled mechanical state produced by the parallel configuration.

\section{\label{sec:Entanglement_verification}Entanglement Verification}
In order to verify the mechanical-mechanical entanglement generated, we propose an experimental procedure in order to measure the components needed to test inseparability criteria for the state produced via the parallel-configuration (Fig. \ref{fig:parallel}). We include the analogous verification set-up for the series configuration in \ref{app:config2}. The verification protocol is tailored towards assessing inseparability criteria belonging to the class of criteria in Ref.~\cite{Shchukin2005}. These inseparability tests take the form of inequalities (constructed from physical observables) which when violated verify the existence of entanglement. The experimental procedure we propose here can provide access to all the inequalities belonging to this class of inseparability criteria. To demonstrate the strength of our scheme we have selected two such inequalities from this class; one of which is guaranteed to detect entanglement in Gaussian states, while the other is able to detect entanglement in states with a greater degree of non-Gaussianity. 

\subsection{\label{sec:inseparability_criteria}Inseparability Criteria}
Numerous inseparability criteria have been proposed to study continuous-variable bipartite states, including operational criteria for Gaussian entanglement detection~\cite{Simon2000, Duan2000, Giedke2001}, and methods for finding optimal continuous-variable entanglement witnesses \cite{Lewenstein2000, Guhne2006, Hyllus2006}. For a review see Ref.~\cite{Adesso2007}. While the von Neumann entanglement entropy quantifies exactly the bipartite entanglement of pure states, it does not have an operational interpretation when the initial mechanical state $\rho_\mathrm{in}$ is a mixed ensemble, e.g. a thermal state. Indeed, many continuous-variable entanglement tests are necessary and sufficient for detecting entanglement in Gaussian states, however not all are capable of capturing entanglement in highly non-Gaussian states (to our knowledge there is no single test that is guaranteed to detect all forms of non-Gaussian entanglement in a state). While Gaussian states are fully characterised by their first and second moments, the same is not true for a non-Gaussian state \cite{serafini_book}. Since higher-order moments are needed to characterise non-Gaussian states, criteria formed from higher-order moments are a natural choice for confirming non-Gaussian entanglement. Shchukin and Vogel introduced a class of inseparability criteria derived from the negative partial transpose (NPT) of the state which take the form of inequalities constructed from arbitrary moments of a continuous-variable quantum state~\cite{Shchukin2005}. Therefore, this class of operational criteria lends itself well to the identification of entanglement in non-Gaussian states. 

The NPT criterion is a sufficient condition for the entanglement of a quantum state~\cite{Peres1996,Horodecki1996}, and for continuous-variable bipartite systems, the criterion manifests itself in the negativity of sub-determinants of a matrix constructed in a specific way from observable moments of the state \cite{Shchukin2005}. The determinant calculated from the first $N$ rows and columns of the matrix is denoted as $D_N$. If $D_N<0$ for any $N$ then NPT has been demonstrated and the state is entangled.
The first few rows and columns of the matrix from which $D_N$ is calculated are shown here:
\begin{equation}
\label{eq:D5}
D_N=\begin{vmatrix}
1 & \braket{{b}_1} & \braket{{b}_1^\dag} & \braket{{b}_2^\dag} & \braket{{b}_2} & \dots\\
\braket{{b}_1^\dag} & \braket{{b}_1^\dag b_1} & \braket{{b}_1^{\dag 2}} & \braket{{b}_1^\dag b_2^\dag} & \braket{{b}_1^\dag b_2} & \dots\\
\braket{{b}_1} & \braket{{b}_1^2} & \braket{{b}_1{b}_1^\dag} &\braket{{b}_1{b}_2^\dag} & \braket{{b}_1{b}_2} & \dots\\
\braket{{b}_2} & \braket{{b}_1{b}_2} & \braket{{b}_1^\dag{b}_2} & \braket{{b}_2^\dag{b}_2} & \braket{{b}_2^2} & \dots\\
\braket{{b}_2^\dag} & \braket{{b}_1{b}_2^\dag} & \braket{{b}_1^\dag{b}_2^\dag} & \braket{{b}_2^{\dag 2}} & \braket{{b}_2 {b}_2^\dag} & \dots\\
\vdots & \vdots	& \vdots & \vdots	&\vdots &\ddots
\end{vmatrix}~.
\end{equation}
Other entanglement criteria are found to exist within this formalism \cite{Duan2000, Mancini2002, Agarwal2005, Raymer2003}. In particular, $D_5$ (the determinant of the first 5 rows and columns of the matrix in Eq. \eqref{eq:D5}) is a reformulation of Simon's criterion \cite{Simon2000} which is a necessary and sufficient entanglement test for single-mode bipartite Gaussian states \cite{Werner2001}. Therefore, applying $D_5$ to a Gaussian state will always verify entanglement (although some non-Gaussian entangled states might also satisfy $D_5<0$ since a negative $D_5$ is only a sufficient condition for non-Gaussian entangled states). 

We can construct other inseparability criteria by deleting rows and columns of the matrix in Eq. \eqref{eq:D5} in a pairwise fashion. If the determinant of the resulting matrix is negative then this also fulfils the NPT criterion and the state is entangled. 
In this way, we can delete entries in $D_N$ to arrive at the following subdeterminant:

\begin{equation}
\label{eq:S3}
S_3=\begin{vmatrix}
1 & \braket{{b}_2^\dag} & \braket{{b}_1{b}_2^\dag}\\
\braket{{b}_2} & \braket{{b}_2^\dag{b}_2} & \braket{{b}_1{b}_2^\dag{b}_2}\\
\braket{{b}_1^\dag{b}_2} & \braket{{b}^\dag_1{b}_2^\dag{b}_2} & \braket{{b}_1^\dag {b}_1 {b}_2^\dag{b}_2}
\end{vmatrix}~,
\end{equation}
such that if $S_3<0$ we have a sufficient criterion for entanglement. $S_3$ is one of the simplest subdeterminants that goes beyond combinations of quadratic moments, thus allowing non-Gaussian entanglement to be detected. As a result of its simplicity, it has been previously highlighted in the context of two-mode Schr{\"o}dinger-cat states~\cite{Shchukin2005,Miranowicz2009}. We note that lower-dimensional matrices, from which the determinant is computed, offer a more practical route to experimental verification. 

Since $D_5$ captures all entanglement for Gaussian states, if $S_3$ indicates entanglement in a region of state space for which $D_5$ does not, we can conclude that these entangled states are non-Gaussian. Therefore, applying these two inequalities together to a state, we can identify parameter regions where the state is non-Gaussian and entangled. We note that there may be entangled non-Gaussian states that are not detected by $D_5$ or $S_3$, since these are only sufficient entanglement criteria. Furthermore, for states where both $S_3<0$ and $D_5<0$, or $D_5<0$ but $S_3>0$, we cannot infer if the entangled state is non-Gaussian.
It should also be highlighted that the magnitude of a determinant has no relevance to the NPT criterion; only the sign of a determinant matters for entanglement verification. Nevertheless, a more negative determinant could be easier to experimentally confirm since each expectation value has an associated experimental uncertainty. We will explore the effect of errors arising from environment on the expectation values in Section~\ref{sec:decoherence_model}.

\subsection{\label{sec:verification_protocol}Verification protocol} 
We now detail our experimental scheme to obtain the moments of the mechanical state which are needed for the inseparability criteria in Section~\ref{sec:inseparability_criteria}. 
Our verification protocol can be used to extract arbitrary mechanical moments. These can then be used to calculate $D_5$ and $S_3$ or more complicated subdeterminants composed of higher-order moments.

\begin{figure*}
    \includegraphics[width=0.9\textwidth]{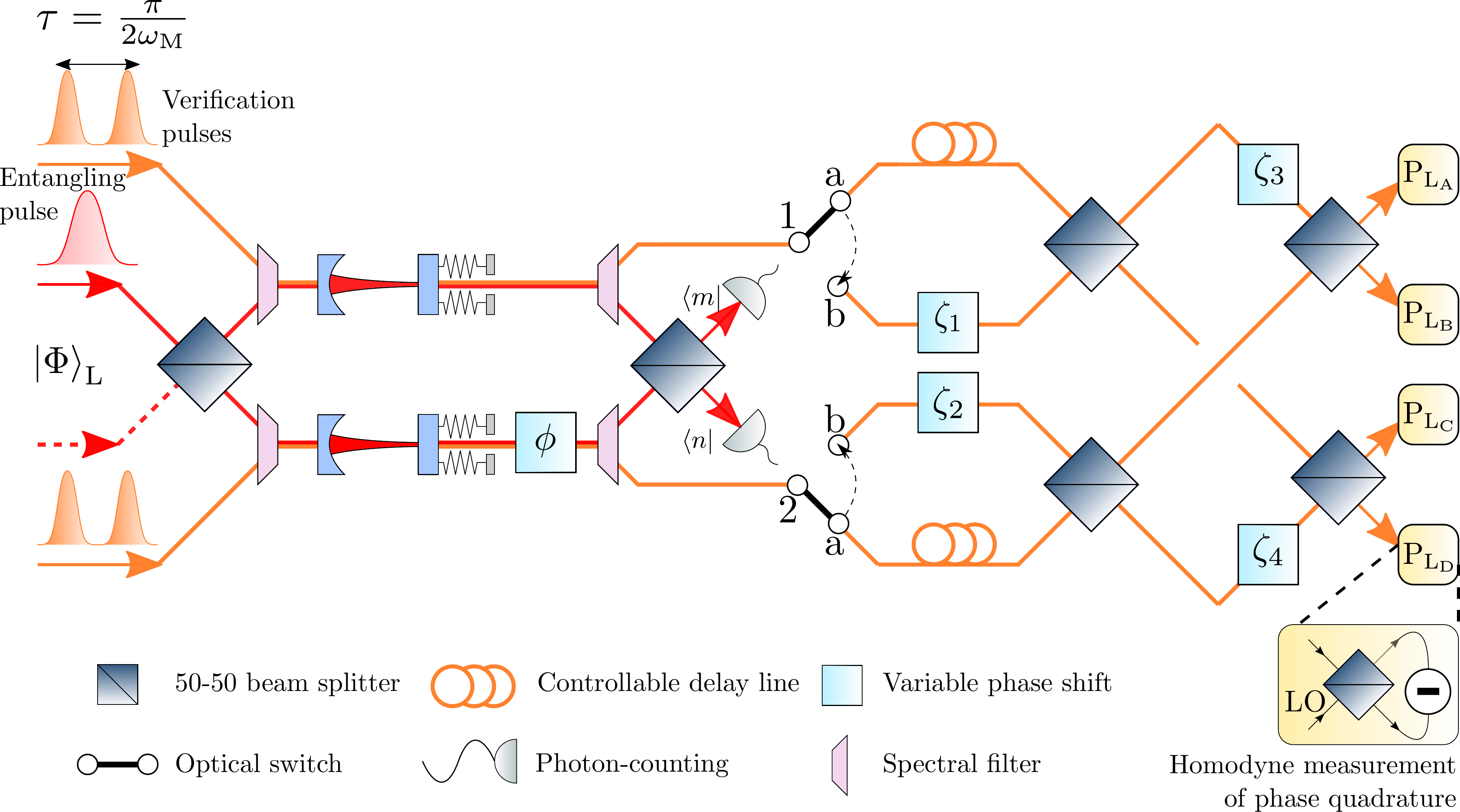}
    \caption{Proposed verification scheme to measure arbitrary quadrature moments of the mechanical state produced in the parallel configuration. These moments allow us to compute the inseparability criteria $D_5$ (Eq. \eqref{eq:D5}) and $S_3$ (Eq. \eqref{eq:S3}). The entangling pulse, indicated by the red path on the diagram, generates a mechanical state. Following a \{1,0\} click event, the verification stage is subsequently conducted. Up to four pulses of light are sent into apparatus (two in each arm) at times $\tau=0$ and/or $\tau=\pi/(2\omega_\mathrm{M})$. These verification pulses, denoted by the orange path on the diagram, have a different wavelength to the entangling pulse and so spectral filters ensure that the verification and entangling pulses follow different optical paths. Both the entangling and verification pulses are on resonance with a cavity mode that interacts in the unresolved sideband regime. Following an optomechanical interation, the light quadratures of the verification pulse transform according to Eqs. \eqref{eq:transform1}-\eqref{eq:transform2}; mechanical quadrature information is transferred to the momentum quadrature of light. Switches 1 and 2 provide the option to hold earlier pulses in a delay line, ensuring all pulses are incident at the beam splitters concurrently. Pulses which do not require a delay follow path 1b or 2b. Variable phase shifts $\zeta_{l}$ (where $l=1,2,3,4$) can be changed between pulses and runs, providing us with the degrees of freedom necessary to solve for the desired moments. Homodyne measurements $P_{\mathrm{L}_k}$ (where $k=A,B,C,D$) are performed on the momentum quadrature of light as it emerges from the final set of beam splitters. For each configuration of pulses, switches, and phase shifts, we require many runs (each requiring state creation followed by a measurement of $P_{\mathrm{L}_k}$) in order to construct a data set of measurements $\{P_{\mathrm{L}_k}^{(i)}\}_{i=1}^N$. From this data set one can calculate moments such as $\braket{P_{\mathrm{L}_k}^d}$ using Eq. \eqref{eq:direct_sum} which can in turn be expanded in terms of mechanical quadrature moments, thus allowing us to find all moments of the form $\braket{X_{\mathrm{M}_1}^p P_{\mathrm{M}_1}^q X_{\mathrm{M}_2}^r P_{\mathrm{M}_2}^s}$.}
    \label{fig:verification_setup}
\end{figure*}

The experimental schematic for the verification protocol is illustrated in Fig. \ref{fig:verification_setup}.
After the entangling pulse has interacted with the two oscillators and photon-detection has been performed, we are left with a mechanical-mechanical entangled state. 
After this photon measurement, up to four verification pulses are sent into each arm of the apparatus. For the verification pulses to provide an independent readout, and to operate in the linearized regime, these fields are resonant with a different optical mode. The pulses also address each oscillator individually further improving the independence of the readout. Assuming the verification pulses are strong and the optical phase shift imparted to the light is small, we get the following linearized optomechanical interaction for each pulse: $U_\mathrm{v}=\mathrm{e}^{\mathrm{i}\chi X_{\textsc{L}_i} X_{\textsc{M}_j}}$, where $X_{\textsc{L}_i}$ and $X_{\textsc{M}_j}$ are the position operators of the $i^{\mathrm{th}}$ light mode and $j^{\mathrm{th}}$ oscillator respectively~\cite{Clarke2020, Brunelli2020, Neveu2021}. We emphasise that the verification pulses are independent from the generation pulse, but to avoid overloading the notation we simply use $\mathrm{L}_i$ in this section to refer to the verification light modes. The dimensionless interaction strength is given by $\chi\propto g_0 \sqrt{N_{\mathrm{p}}}/\kappa$ where $N_{\mathrm{p}}$ is the mean photon number in the input field. The coupling rate here $g_0$ differs to the coupling rate for the state-preparation stage as a different cavity mode is used, and the proportionality constant is of order unity, which depends on the properties of the input pulse. If we use a long verification pulse with duration $\tau$ that satisfies $\omega_{\mathrm{M}}\ll \tau ^{-1}\ll \kappa$, then $\chi = 4 g_0 \sqrt{N_{\mathrm{p}}}/\kappa$~\cite{khosla2013quantum,Bennett2016}. Similar to the generation stage, using such a pulse has the advantage of more easily ensuring spatio-temporal mode matching at the second beam splitter if the two cavities are not identical. Note also that an additional deterministic momentum kick is also imparted on the system per verification pulse that depends on the optical intensity, which may be accounted for in post-processing or cancelled with an appropriate feedback force.

Following the interaction $U_{\mathrm{v}}$, the $i^{\mathrm{th}}$-mode light quadratures $X_{\textsc{L}_i}$ and $P_{\textsc{L}_i}$ are given by: 
\begin{subequations}
    \begin{align}
        X_{\textsc{L}_i}(\theta_j)&=X_{\mathrm{L}_\mathrm{in}}~,\label{eq:transform1}\\
        P_{\textsc{L}_i}(\theta_j)&=P_{\mathrm{L}_\mathrm{in}}+\chi X_{\textsc{M}_j}(\theta_j)~,\label{eq:transform2}
    \end{align}
\end{subequations}
where $X_{\textsc{M}_j}(\theta_j)$ is the mechanical position operator before the interaction $U_{\mathrm{v}}$. Here, we have assumed the input light quadratures $X_{\mathrm{L}_\mathrm{in}}$ and $P_{\mathrm{L}_\mathrm{in}}$ are quantum noise limited with variances of 1/2, and the mechanical quadratures freely evolve as a function of $\theta_j=\omega_{\textsc{M}_j}\tau$, where $\omega_{\textsc{M}_j}$ is the mechanical frequency of the $j^{\mathrm{th}}$ oscillator, and $\tau$ is the time elapsed between state generation and the interaction $U_{\mathrm{v}}$. We henceforth assume that the two oscillators have the same parameters for brevity and drop the $j$ label on $\omega_\mathrm{M}$. For a system isolated from its environment, $X_{\textsc{M}_j}(\pi/2)=P_{\textsc{M}_j}(0)$, so by switching on the verification pulse at different times we can imprint different mechanical quadratures on $P_{\textsc{L}_{i}}$. 

As $X_\mathrm{M}, P_\mathrm{M}$ are more experimentally accessible than $b$, $b^\dag$ we will recast the determinants in Eqs. \eqref{eq:D5} and \eqref{eq:S3} to be in terms of these operators using the relation $b_i=(X_{\mathrm{M}_i}+\mathrm{i}P_{\mathrm{M}_i})/\sqrt{2}$. From this, each element of $D_N$ (Eq. \eqref{eq:D5}) can be expressed as a linear combination of moments in $X_{\mathrm{M}}$ and $P_{\mathrm{M}}$: $\sum_{pqrs} c_{pqrs}\braket{X_{\mathrm{M}_1}^p P_{\mathrm{M}_1}^q X_{\mathrm{M}_2}^r P_{\mathrm{M}_2}^s}$ where $c_{pqrs}\in\mathbb{C}$.
This is convenient as the optomechanical interaction in Eq. \eqref{eq:transform2} is dependent on the mechanical quadratures.

To obtain all the terms in the inseparability criteria we must measure moments of the form $\braket{X_{\mathrm{M}_1}^p P_{\mathrm{M}_1}^q X_{\mathrm{M}_2}^r P_{\mathrm{M}_2}^s}$. Since we cannot simultaneously measure $X_{\mathrm{M}_i}$ and $P_{\mathrm{M}_i}$, we exploit a network of switches, time delays, and phase shifts $\{\zeta_l\}$ in the optical set-up (Fig. \ref{fig:verification_setup}) to probe these quantities. This network should be integrated with the optical apparatus used for entanglement generation as each verification run involves recreating the entangled state. Our proposed scheme allows us to control which combination of mechanical quadratures are obtained via homodyne measurement of the output light quadratures $P_{\mathrm{L}_k}$, where $k=\{A,B,C,D\}$ (see Fig. \ref{fig:verification_setup}). 

We can build a data set of $N$ homodyne measurement results $\{P_{\mathrm{L}_k}^{(i)}\}_{i=1}^N$ for a particular optical pathway (a specific combination of switches, time-delays, and phase shifts characterises a pathway) by repeatedly generating the mechanical state, and then performing a verification pulse with homodyne detection. From the data we can directly compute moments using the formula:
\begin{equation}
\label{eq:direct_sum}
    \braket{P_{\mathrm{L}_k}^d}=\frac{1}{N}\sum_i^{N}(P^{(i)}_{\mathrm{L}_k})^d~.
\end{equation}
Eq. \eqref{eq:transform2} allows us to expand the measured $\braket{P_{\mathrm{L}_k}^d}$ as a linear combination of mechanical quadrature moments with coefficients determined by the phase shifts $\{\zeta_l\}$. Repeating these steps with sufficiently many appropriate combinations of $\{\zeta_1,\zeta_2,\zeta_3,\zeta_4\}$ will provide us with enough linearly independent equations to solve for any of the mechanical quadrature moments that appear in the expansion of $\braket{P_{\mathrm{L}_k}^d}$~\cite{Opatrny1997}. From these moments $D_5$, $S_3$, or any other subdeterminant can then be calculated. 

As a simple example we can consider finding $\braket{X_{\mathrm{M}_1}}$. After a \{1,0\} click heralds the state $\rho^{\{1,0\}}_{\mathrm{out}}$, the verification pulses are allowed to interact with the mechanical oscillators. Spectral filters ensure the verification pulses follow a different pathway to that of the entangling pulse, since the entangling and verification pulses have a different wavelength (see Fig. \ref{fig:verification_setup}). For this mechanical quadrature moment, a single verification pulse is injected at time $\tau=0$ into the upper arm and interacts with oscillator 1. The pulse then follows path 1b, passes through a set of beam splitters (for this moment $\{\zeta_{l}\}$ are unimportant). Finally, we perform a homodyne measurement of $P_{\mathrm{L}_k}$ (Eq. \eqref{eq:transform2}). Over numerous runs (each time recreating the entangled state) we build a data set of $N$ homodyne measurements $\{P_{\mathrm{L}_k}^{(i)}\}_{i=1}^N$. According to Eq. \eqref{eq:direct_sum}, by summing over the data set we can deduce the first moment $\braket{P_{\mathrm{L}_k}}=\chi\braket{X_{\mathrm{M}_1}}/2$ (where we have assumed vacuum noise statistics $\braket{P_{\mathrm{L}_\mathrm{in}}}=0$, and the factor $1/2$ arises from the beam splitters). The value of $\chi$ is assumed to be known accurately via an initial calibration stage (see \ref{app:calibration}), allowing us to extract the value of $\braket{X_{\mathrm{M}_1}}$. Repeating this entire sequence but sending in a single pulse at time $\tau=\pi/(2\omega_{\textsc{M}})$ will yield $\braket{P_{\mathrm{M}_1}}$. Using Eq. \eqref{eq:direct_sum} we can calculate higher-order moments of $P_{\mathrm{L}_k}$ from the same data set and thus find higher order mechanical moments. For example, the third moment can be expressed as $\braket{P^3_{\mathrm{L}_k}}=\chi^3\braket{X^3_{\mathrm{M}_1}}/8+\chi\braket{X_{\mathrm{M}_1}}/4$ (where we have assumed $\braket{P^3_{\mathrm{L}_{\mathrm{in}}}}=0$ and $\braket{P^2_{\mathrm{L}_{\mathrm{in}}}}=1/2$). This contains the already-calculated term $\braket{X_{\mathrm{M}_1}}$, and $\braket{P_{\mathrm{L}_k}^3}$ is found from the data set of homodyne measurements; and so we can now determine $\braket{X^3_{\mathrm{M}_1}}$. Continuing this iterative procedure, we can use lower-order moments to iteratively find those of higher order. 

The full method for finding an arbitrary mechanical moment $\braket{X_{\mathrm{M}_1}^pP_{\mathrm{M}_1}^qX_{\mathrm{M}_2}^rP_{\mathrm{M}_2}^s}$ is outlined in \ref{app:general}, while a simpler experimental approach for finding some of the lower-order moments in $D_5$ and $S_3$ is presented in \ref{app:special}. The iterative nature of our verification scheme (using lower-order moments to find higher ones) means that, by determining all the moments for $S_3$, one also unlocks all the lower-order moments for $D_5$, making it experimentally convenient to calculate both determinants together. We note that, in practice, to confidently measure higher-order moments using Eq. \eqref{eq:direct_sum} and thus unambiguously determine the sign of the determinants $D_5$ and $S_3$, we must ensure a sufficiently large data set of homodyne measurement results $\{P_{\mathrm{L}_k}^{(i)}\}_{i=1}^N$ for each optical pathway considered. We note that the standard error on $\braket{P_{\mathrm{L}_k}^d}$ is $\sigma/\sqrt{N}$, where $\sigma$ is the standard deviation of $\braket{P_{\mathrm{L}_k}^d}$ which can be experimentally calculated using the data set. Thus increasing $N$ will reduce this error and consequently mitigate the propagation of errors as we iteratively solve for higher-order moments.
Nevertheless, unlike pure optical schemes \cite{Shchukin2005PhysRevA.72.043808}, our protocol allows increasingly high-order moments of the bipartite state to be found without requiring additional optical apparatus (only those in Fig. \ref{fig:verification_setup}). 

\subsection{\label{sec:decoherence_model}Open-system dynamics}
In the verification protocol described in the previous section, we assumed that the mechanical state is isolated from its environment while the verification stage is performed. This idealised version captures the essence of the scheme; however, we now consider the effects that occur from the system interacting with a thermal environment during the time elapsed between state preparation and the verification stage. We henceforth assume that the two oscillators are in identical thermal environments, and share the same mechanical properties and initial thermal state ($\bar{n}_1=\bar{n}_2=\bar{n}$). The mechanical quadratures evolve between $X_{\textsc{M}_i}$ and $P_{\textsc{M}_i}$ and undergo damping and rethermalisation due to their coupling with an external heat bath. This behaviour is described by the following quantum Langevin equations:
\begin{subequations}
    \begin{align}
    \dot{X}_{\mathrm{M}_i}&=\omega_{\mathrm{M}} P_{\mathrm{M}_i}~, \label{eq:Langevin_X} \\
    \dot{P}_{\mathrm{M}_i}&=-\omega_{\mathrm{M}} X_{\mathrm{M}_i} -\gamma P_{\mathrm{M}_i} +\sqrt{2\gamma}\xi_i~,\label{eq:Langevin_P}
    \end{align}
\end{subequations}
where $\gamma$ is the damping rate, and $\xi_i$ is Brownian force term acting on the $i^{\mathrm{th}}$ oscillator which models random excitations from the bath~\cite{MilburnWoolley}. The Brownian force has the following properties:
\begin{subequations}
\begin{align}
    \braket{\xi_i(t)}&=0~, \label{eq:brownian1}\\
    \braket{\xi_i(t)\xi_j(t')}&=(2\bar{n}_\textsc{B}+1)\delta(t-t')\delta_{ij}~,\label{eq:brownian2}
\end{align}
\end{subequations}
where $\bar{n}_\textsc{B}$ is the thermal occupation number of the bath. In general, $\bar{n}\neq\bar{n}_{\mathrm{B}}$ if cooling strategies are implemented. The time-dependent solutions to these coupled differential equations are given by Eqs. \eqref{eq:Langevin_solution_X}-\eqref{eq:Langevin_solution_P} in \ref{app:decoherence}. The influence of the environment will manifest during the time delays between verification pulses and we characterise this effect through $\bar{n}_{\mathrm{B}}$, and the oscillators' quality factors $Q=\omega_\mathrm{M}/\gamma$.

\subsection{Single-photon detector considerations}\label{sec:experimental_considerations}
With a single photon input state, the use of a $\{1,0\}$ click event to herald the mechanical state $\rho^{\{1,0\}}_{\mathrm{out}}$ is robust against optical losses: if a photon is lost in the interferometer, no click is detected and this run of the experiment is discarded. In contrast, a coherent state input risks creating a mechanical state which is different to that indicated by the photodetector clicks. In Section~\ref{sec:Entanglement_Protocol} we noted that when $\ket{\Phi}_{\mathrm{L}}=\ket{\alpha}_1\ket{0}_2$, choosing $|\alpha|=1$ maximises the probability of heralding the state $\rho^{\{1,0\}}_{\mathrm{out}}$ (Eq. \eqref{eq:heralding_prob}). However, it is possible that a different mechanical state $\rho^{\{m,n\}}_{\mathrm{out}}$ (where $\{m,n\}\neq\{1,0\}$) is created even when the detectors have recorded a $\{1,0\}$ click due to optical losses, detector inefficiencies, or dark counts. We would then incorrectly identify this state as $\rho^{\{1,0\}}_{\mathrm{out}}$, resulting in a `false positive' event. Conversely, if a \{1,0\} click does successfully herald the creation of $\rho^{\{1,0\}}_{\mathrm{out}}$ this is a `true positive' event. We can model the optical losses and detector inefficiencies by introducing loss-model beam splitters to Figs. \ref{fig:parallel} and \ref{fig:series} with intensity transmission $\eta$ (see Fig. \ref{fig:parallel_setup_loss} in \ref{app:optical_losses}) such that when $\eta=1$ there are no optical losses and the detectors are perfectly efficient. Following the derivation in \ref{app:optical_losses}, we demonstrate that when using number-resolving photodetectors, the fraction $\mathcal{F}$ of true positive events per total number of \{1,0\} clicks is expected to be:
\begin{equation}\label{eq:false_positives_ratio_resolving}
    \mathcal{F}=\left[e^{(1-\eta)|\alpha|^2}+\frac{e^{-\eta|\alpha|^2}\mathcal{D}}{\eta {P}_{10}(1-\mathcal{D})}\right]^{-1}~,
\end{equation}
where $\mathcal{D}$ is the probability of detecting a single dark count in the detection window, and ${P}_{10}$ is defined in Eq. \eqref{eq:heralding_prob} (for $\ket{\Phi}_{\mathrm{L}}=\ket{\alpha}_1\ket{0}_2$). The pulsed regime constrains the detection window to a small enough duration (10~ns considered here using parameters discussed below) such that we can assume $\mathcal{D}$ is on the order of $10^{-8}$, and that the probability of multiple dark counts in the window is negligible. In order to neglect false positives and confidently ensure that a $\{1,0\}$ click heralds the $\rho_{\mathrm{out}}^{\{1,0\}}$ state, $\mathcal{F}$ must be as close to unity as possible. Therefore, we require that $(1-\eta)|\alpha|^2 \ll 1$, and that $\mathcal{D}$ is sufficiently low compared to ${P}_{10}$. Thus, Eqs \eqref{eq:heralding_prob} and \eqref{eq:false_positives_ratio_resolving} can be used to establish the feasibility for the heralding process, in particular, for the dependence on $\mu$.

We can derive a similar condition to Eq.~\eqref{eq:false_positives_ratio_resolving} for the case of non-resolving detectors which is given by Eq. \eqref{eq:false_positives_ratio_non_resolving} in \ref{app:optical_losses}. The non-resolving approach similarly requires $\mathcal{D}$ and optical loss to be sufficiently low such that we can neglect false positives. The behaviour of this ratio is further examined in Section~\ref{sec:discussion_optical}.

\section{\label{sec:results}Results}
\begin{table*}[t!]%[htbp]
  \caption{Parameter sets of proposed and experimentally-realised values; considered for the parallel entanglement configuration with $\phi=\pi$, where $\phi$ is the phase in Fig. \ref{fig:parallel}. The two oscillators are characterised by the dimensionless optomechanical coupling strength $\mu$, their quality factor $Q$, their initial thermal occupation number $\bar{n}$, and the occupation of the thermal bath $\bar{n}_\mathrm{B}$. 
  The inseparability criteria $D_5$ and $S_3$ have been calculated to incorporate the open-system dynamics which the system undergoes in the time elapsed between state generation and verification (see Section~\ref{sec:decoherence_model}). 
  We assume the following throughout: $\eta=0.8$ (overall optical intensity efficiency), $\mathcal{D}=10^{-8}$ (the probability of a single dark count during the detection window; assuming a dark count rate of 1~s$^{-1}$ and a detection window of 10~ns). 
  The fraction $\mathcal{F}$ of true positive \{1,0\} events is shown both for the case of number-resolving photodetectors (Eq. \eqref{eq:false_positives_ratio_resolving}) and non-resolving (Eq. \eqref{eq:false_positives_ratio_non_resolving}).
  The main values of $\mathcal{F}$ are calculated with $\alpha=1$ (the amplitude of the injected coherent state in the entanglement stage) and the values in brackets show $\mathcal{F}$ maximised by tuning $\alpha$.
  (i)-(iv) are theoretically proposed parameters. Experiment-inspired parameter sets are used based on Refs \cite{Rossi2018}, a micro-mechanical membrane; \cite{Leijssen2017}, a nanobeam oscillator; and \cite{Wilson2015}, a sliced photonic-crystal structure. The rows corresponding to each of these three references are divided into two sub-rows: the first sub-row contains published protocol parameters, in the second-row we use values for $\bar{n}$ and $\bar{n}_\mathrm{B}$ based on near-future improvements (involving additional cooling techniques) of existing systems. For a pulse which satisfies $\omega_{\mathrm{M}}\ll \tau^{-1}\ll \kappa$, it follows that $\mu=2\sqrt{2}g_0/\kappa$ (for example experimental parameters see Table \ref{tab:extra_parameters}). $^*$These $\bar{n}_\mathrm{B}$ values have been calculated assuming the thermal bath has been cooled to 100~mK.}
\makebox[\textwidth][c]{
    %\begin{ruledtabular}
    \begin{tabular}{c|cccc|cc|cc}
    \hline\hline
& \multicolumn{4}{c|}{Protocol parameters} & \multicolumn{2}{c|}{\textbf{Determinants}} & \multicolumn{2}{c}{True positives $\mathcal{F}~ [\%]$} \\
\midrule
Refs. & $\mu$ & $Q$ & $\bar{n}$  & $\bar{n}_\mathrm{B}$ & $\mathbf{D_5}$ & $\mathbf{S_3}$ & Resolving & Non-resolving \\ \hline 
(i) & $10^{-3}$ & $10^{5}$ & 0.1 & 1000 &  0.56 & $-0.080$ & $79\%~(84\%)$ & $79\%~(84\%)$\\
(ii) & $10^{-2}$ & $10^{5}$ & 0.1 & $1000$ & 0.56 & $-0.080$  & $82\%~(98\%)$ & $82\%~(98\%)$ \\
(iii) & $10^{-1}$ & $10^5$ & 0.1 & 1000 & 0.56 & $-0.080$ & $82\%~(>99\%)$ & $82\%~(>99\%)$\\
(iv) & $10^{0}$ & $10^5$ & 0.1 & 1000 & 0.54 & $-0.029$ & $82\%~(>99\%)$ & $66\%~(>99\%)$ \\ \hline
\cite{Rossi2018} & $2.26\!\times\!10^{-5}$ & $1.03 \!\times\! 10^{9}$ & $0.29$ & $10^5$ & $1.4$ & $0.084$ & $1.3\%~(2.8\%)$ & $1.3\%~(2.8\%)$ \\ 
 & $2.26 \!\times\! 10^{-5}$ & $1.03 \!\times\! 10^{9}$ & $0.1$ & $10{^3}^*$ & $0.51$ & $-0.089$ & $0.99\%~(2.1\%)$ & $0.99\%~(2.1\%)$\\ \hline
\cite{Wilson2015} & $1.29 \!\times\! 10^{-4}$ & $7.54 \!\times\! 10^5$ & $5.3$ & $2.1\!\times\!10^{4}$ & $3500$ & $420$ & $65\%~(65\%)$ & $65\%~(65\%)$\\
& $1.29 \!\times\! 10^{-4}$ & $7.54 \!\times\! 10^5$ & $0.1$ & $484^*$ & $0.51$ & $-0.089$ & $23\%~(30\%)$ & $23\%~(30\%)$\\ \hline
\cite{Leijssen2017} & $1.12 \!\times\! 10^{-2}$ & $3.74 \!\times\! 10^4$ & $1.7 \!\times\! 10^{4}$ & $1.7\!\times\! 10^{4}$ & $1.3\!\times\! 10^{17}$ & $6.3\!\times\! 10^{12}$ & $82\%~(>99\%)$  & $60\%~(>99\%)$ \\
& $1.12 \!\times\! 10^{-2}$ & $3.74 \!\times\! 10^4$ & 0.1 & $559^*$ & 0.58 & $-0.074$ & $82\%~(98\%)$ & $82\%~(98\%)$\\
\hline\hline
\end{tabular}%
%\end{ruledtabular}
}
\label{tab:proposed_parameters}%
\end{table*}%

\floatsetup[figure]{style=plain,subcapbesideposition=top}
\begin{figure*}[]
\centering
\sidesubfloat[]{%
  \includegraphics[height=6cm]{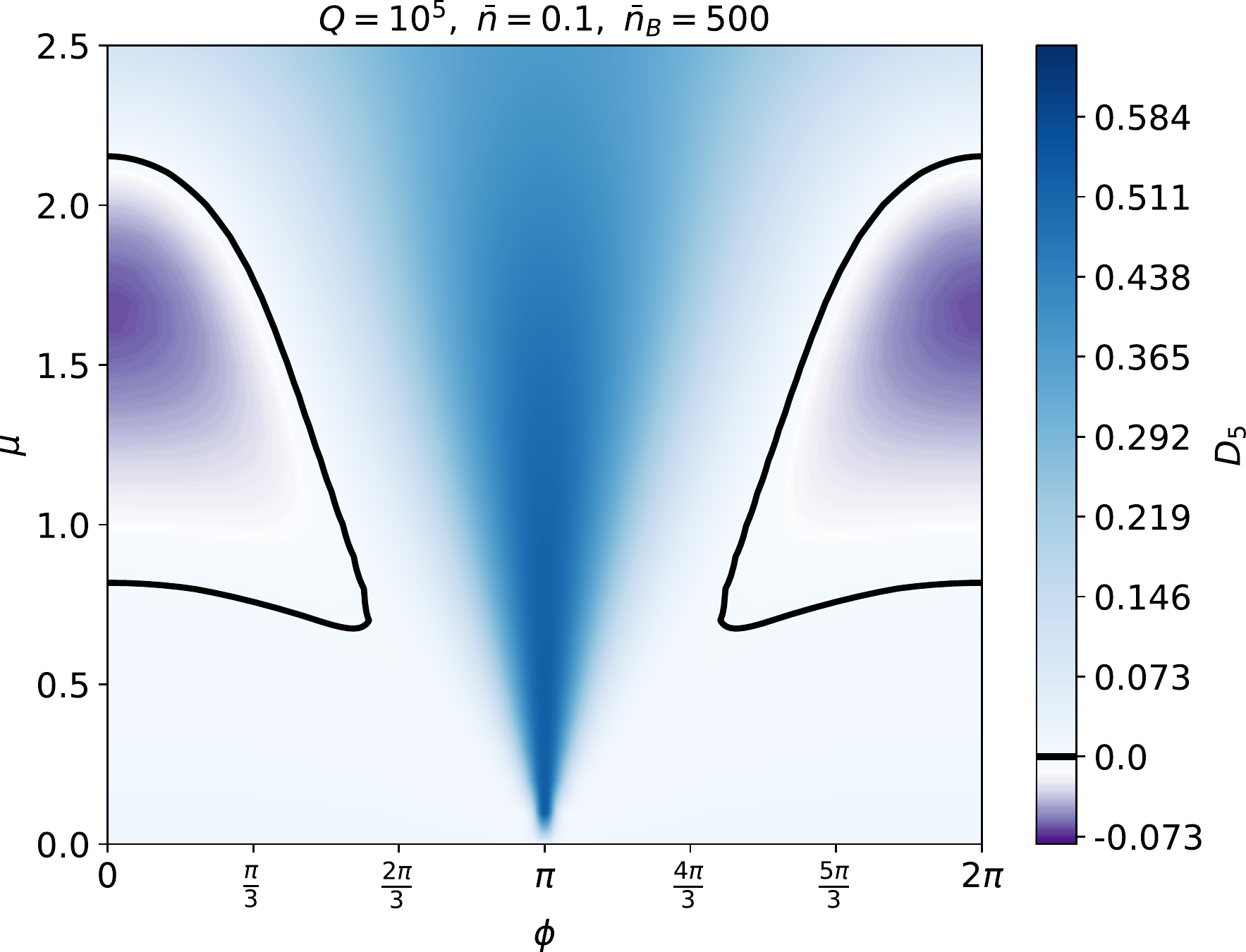}%
  \label{fig:D5}
} \qquad
\sidesubfloat[]{%
  \includegraphics[height=6cm, ]{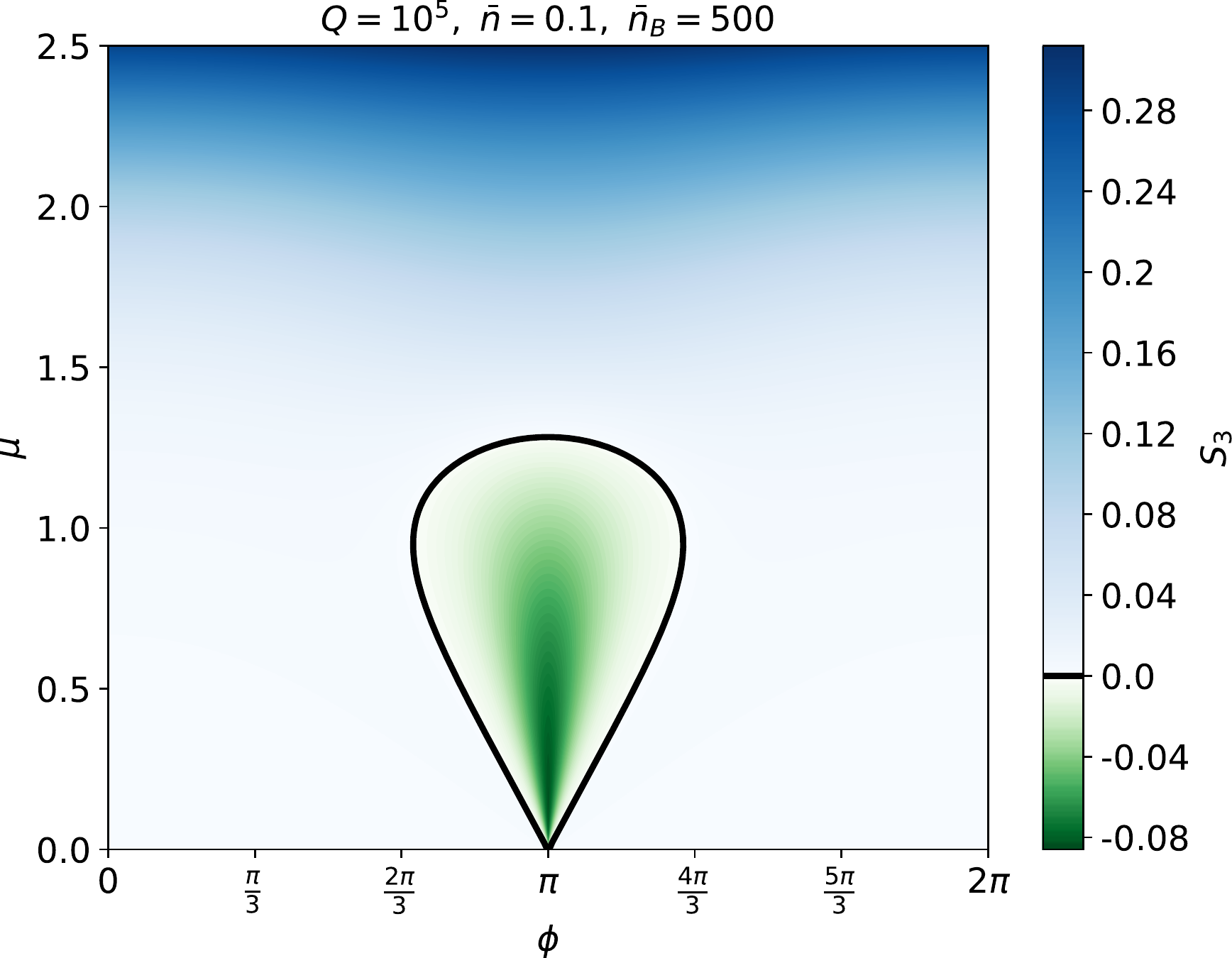}
  \label{fig:S3}
}\hfill
\caption{Inseparability criteria outcomes as obtained via the verification method for the state $\rho^{\{1,0\}}_{\mathrm{out}}$ produced via the parallel configuration (Fig. \ref{fig:parallel}). Plots of $D_5$ (a) and $S_3$ (b) as functions of the interferometer phase $\phi$, and the optomechanical coupling strength $\mu$. The open-system dynamics predicted to occur while measuring the determinants (see Fig. \ref{fig:verification_setup}) is parameterised by $Q=10^{5}$, $\bar{n}=0.1$, and $\bar{n}_\mathrm{B}=500$; for a summary of the experimental parameters see Table \ref{tab:proposed_parameters}. Both determinants are periodic with $\phi$ and repeat every $2\pi$. The black line demarcates regions where determinants are negative therefore indicating regions where the state is found to be entangled. All Gaussian entangled states fall within the purple region of (a). Non-Gaussian entangled states are captured by the Green region of (b) since it does not overlap with the purple region of (a).}
\end{figure*}

In this section we present our theoretical model results for the verification of entanglement between two optomechanical oscillators which have been prepared in the state $\rho^{\{1,0\}}_{\mathrm{out}}$ using the parallel set-up shown in Fig. \ref{fig:parallel}.
The oscillators are initialized in a separable thermal state $\rho_\mathrm{in}$ with equal $\bar{n}$. For a summary of the key experimental parameters used to calculate the results see Table \ref{tab:proposed_parameters}. For an example experimental parameter set that goes into Table \ref{tab:proposed_parameters} see Table \ref{tab:extra_experimental_parameters} in \ref{app:experimental_params}.

Firstly we present our results for the inseparability criteria $D_5$ (Eq. \eqref{eq:D5}) and $S_3$ (Eq. \eqref{eq:S3}). The moments for these criteria were calculated using the open-system dynamics outlined in Section~\ref{sec:decoherence_model}. This model allows us to simulate the loss of energy to and random excitations from the environment that would arise in an experimental implementation of the verification scheme outlined in Fig. \ref{fig:verification_setup}. As previously mentioned, we assume that both oscillators are initially identical with the same quality factor $Q$, thermal mechanical occupation number $\bar{n}$, and bath occupation number $\bar{n}_\mathrm{B}$. 

Figs. \ref{fig:D5} and \ref{fig:S3} show $D_5$ and $S_3$, respectively, as functions of the experimentally controlled phase $\phi$, and the optomechanical coupling $\mu$. We have chosen $Q=10^5$, $\bar{n}_\mathrm{B}=500$, and $\bar{n}=0.1$. The state $\rho^{\{1,0\}}_{\mathrm{out}}$ has a phase periodicity of $2\pi$ in $\phi$, which is reflected in the determinants. The parameter regions for which entanglement can be verified using $D_5$ and $S_3$ are indicated by the negative values and demarcated by solid black lines. For $D_5$, the negative region is centred at $\phi=\{0,2\pi\}$ while for $S_3$, it is concentrated around $\phi=\pi$. We reiterate that it is the sign of the determinants, not their magnitude which indicates entanglement. Nevertheless, a more negative determinant could be more resistant to experimental errors. That is, for successful verification, the magnitude of a negative determinant should be greater than the uncertainty of the value. 

\begin{figure}[]
         \centering
         \includegraphics[height=6cm]{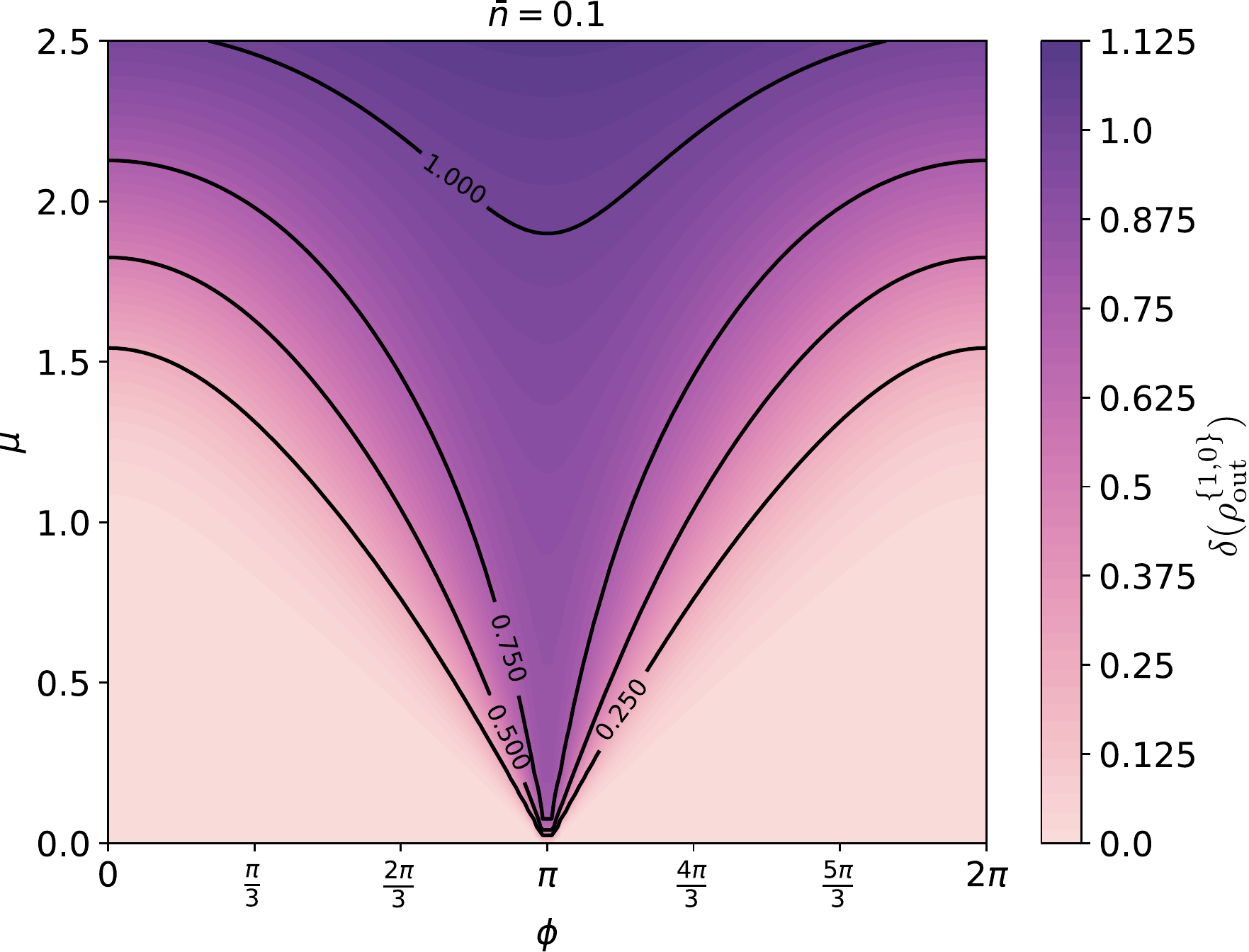}
        \caption{Measure of non-Gaussianity computed on the mechanical state $\rho^{\{1,0\}}_{\mathrm{out}}$ created via the entanglement scheme in Fig. \ref{fig:parallel}. The mechanical oscillators have initial thermal occupation number $\bar{n}=0.1$. The state becomes more non-Gaussian as $\phi\rightarrow\pi$ or as $\mu$ is increased. We note that parameter regions which are more Gaussian qualitatively correlate with the negative contours in Fig. \ref{fig:D5}. Regions which are more non-Gaussian are centred around the same phase ($\phi=\pi$) as the negative contours in Fig \ref{fig:S3}.}
         \label{fig:non_gaussian}
\end{figure}

To better understand the features of Figs. \ref{fig:D5} and \ref{fig:S3} we examine the non-Gaussianity of the state $\rho^{\{1,0\}}_{\mathrm{out}}$ in terms of a non-Gaussian quantifier $\delta(\rho^{\{1,0\}}_{\mathrm{out}})$. This measures the quantum relative entropy between $\rho^{\{1,0\}}_{\mathrm{out}}$ and a Gaussian reference state $\rho_\mathrm{G}$, constructed using the first and second moments of $\rho^{\{1,0\}}_{\mathrm{out}}$. As shown in Ref.~\cite{Genoni2008,Genoni2010}, this measure $\delta$ can be expressed as $\delta(\rho)=S(\rho_\mathrm{G})-S(\rho)$, where $S$ is the standard von Neumann entropy. Since $\delta$ is an exact measure of non-Gaussianity~\cite{Marian2013}, a greater $\delta$ suggests a greater degree of non-Gaussianity, while $\delta=0$ implies our state is exactly Gaussian. However, the measure has no upper bound (there is no maximally non-Gaussian state) so we can only comment on a state being more non-Gaussian than another. 

While we do not propose an experimental way to obtain $\delta$, the measure $\delta(\rho^{\{1,0\}}_{\mathrm{out}})$ is shown in Fig. \ref{fig:non_gaussian} as a function of $\mu$ and $\phi$ for $\bar{n}=0.1$. For simplicity, we have computed the non-Gaussianity of the heralded state before verification has been conducted (i.e. for a closed system). As $\mu\rightarrow 0$, the state becomes more Gaussian; physically at $\mu=0$  there is no optomechanical interaction, therefore the state $\rho^{\{1,0\}}_{\mathrm{out}}$ is comprised of two separable, fully Gaussian thermal states. For a given $\mu$, the state becomes more Gaussian as $\phi\rightarrow 0,2\pi$; likewise, the negative regions of $D_5$ where entanglement can be verified are centred at $\phi=0,2\pi$. This is in line with the established notion that $D_5$ can detect all entanglement in bipartite Gaussian states. In contrast, for a given $\mu$ the state becomes increasingly non-Gaussian as $\phi\rightarrow \pi$. Comparing Figs. \ref{fig:S3} and \ref{fig:non_gaussian}, we see that $S_3$ performs best around the values of $\phi$ that correspond to higher non-Gaussianity. However, in high $\mu$ regimes, we are unable to verify entanglement with $S_3$ despite the state becoming more non-Gaussian as $\mu$ increases (we discuss this feature in Section~\ref{sec:discussion_verification}).

\begin{figure}[]
         \centering
         \includegraphics[height=6cm]{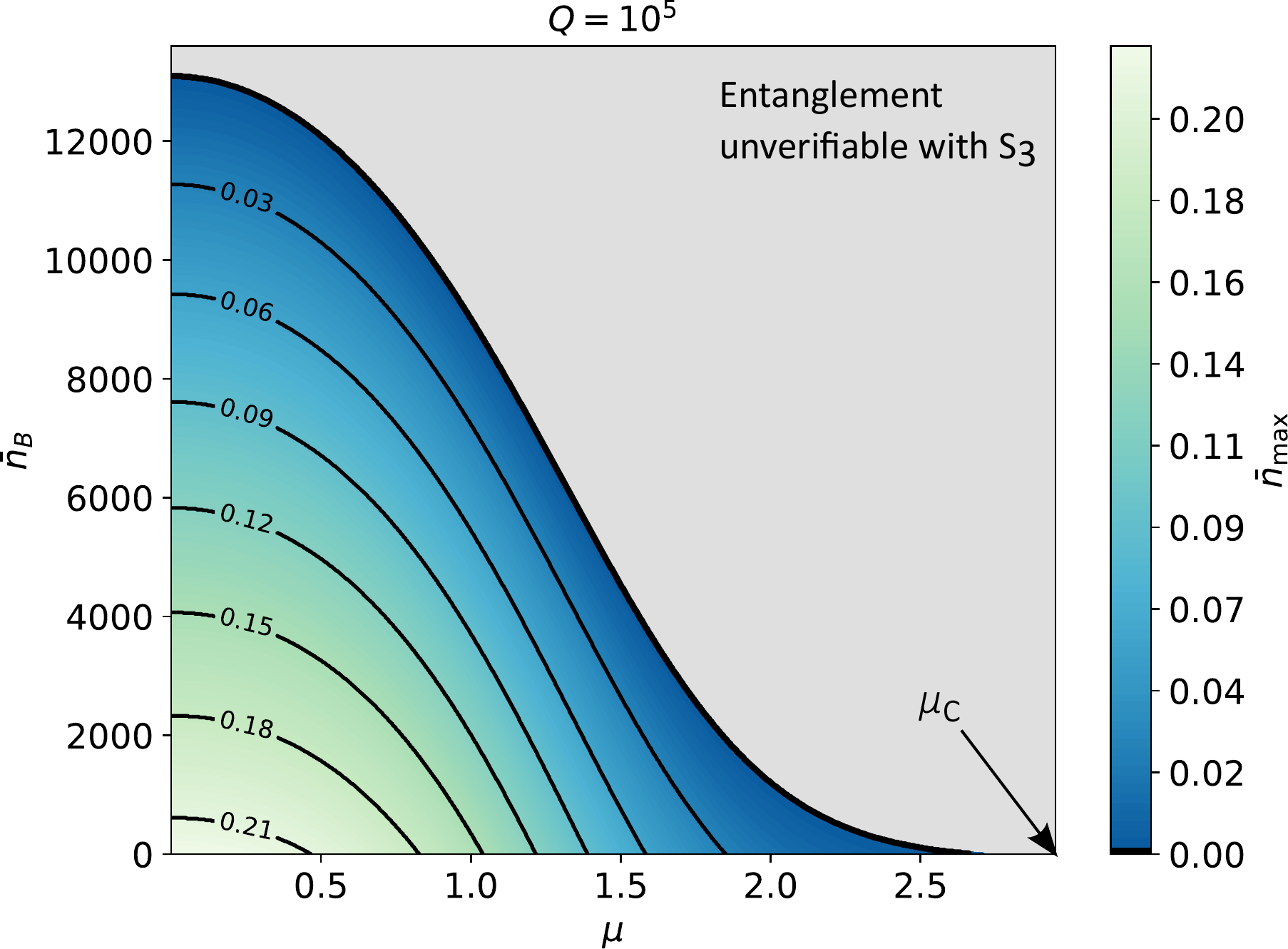}
        \caption{Contour plot of the maximum oscillator thermal occupation $\bar{n}_{\mathrm{max}}$ for which $S_3$ can verify entanglement. This maximum is plotted with $\mu$ and $\bar{n}_\mathrm{B}$ for an oscillator of $Q=10^5$. The phase in the entanglement stage is set to $\phi=\pi$. The gray region corresponds to the parameter space for which $S_3$ is unable to verify entanglement even at $\bar{n}=0$.
        Note, for coupling strengths greater than $\mu_\mathrm{c}$, the scheme is unable to verify entanglement using $S_3$ even at $\bar{n}=\bar{n}_{\mathrm{B}}=0$ due to the interactions with the environment present with finite $\gamma$. 
        }
         \label{fig:n_cooling_interpolated}
\end{figure}

The ability to verify entanglement is strongly dependent on the mechanical initial thermal occupation $\bar{n}$. This can be reduced by lowering the temperature of the oscillator using laser cooling techniques and/or cryogenics. Here we determine the levels of cooling (in terms of $\bar{n}$) required for entanglement to be verified. We are specifically interested in non-Gaussian entanglement and therefore investigate how high we can allow $\bar{n}$ to be while still achieving a negative $S_3$. For this analysis, the phase in the entanglement stage is fixed to $\phi=\pi$ (at which point the mechanical state is most non-Gaussian). For a given $\mu$, $\bar{n}_\mathrm{B}$, and $Q$, the oscillators can take a range of thermal occupation numbers between $0<\bar{n}<\bar{n}_{\mathrm{max}}$ over which $S_3$ is still negative. When $\bar{n}=\bar{n}_{\mathrm{max}}$, $S_3=0$ and we can no longer verify entanglement. The maximum mechanical occupation number $\bar{n}_{\mathrm{max}}$ is shown as a function of $\mu$ and $\bar{n}_\mathrm{B}$ in Fig. \ref{fig:n_cooling_interpolated} for $Q=10^5$. As expected, in order to compensate for an increasing bath temperature one must pre-cool the oscillator more. The gray region in Fig. \ref{fig:n_cooling_interpolated} bounded by the black line designates the values of $\mu$ and $\bar{n}_\mathrm{B}$ for which $S_3$ is not capable of verifying entanglement, even as the oscillator approaches the ground state ($\bar{n}\to0$). We note that beyond a cut-off value of $\mu_\mathrm{c}\approx 3.6$, $S_3$ cannot verify entanglement regardless of bath temperature and oscillator cooling. This does not preclude anything about the entanglement structure of the state at higher $\mu$ but would suggest $S_3$ is more suitable for lower $\mu$ regimes (this behaviour is further discussed in Section~\ref{sec:discussion_verification}). In Table \ref{tab:proposed_parameters}, we give the values of $D_5$ and $S_3$ calculated for seven chosen parameter sets (four theoretically proposed and three taken from state-of-the-art experiments). The experiments that we consider operate in the unresolved sideband regime and so are suitable for our protocol. For the parameter sets taken from these experiments, we include both published parameters and suggested near-future improvements to the oscillator occupation number $\bar{n}$, and the thermal bath occupation number $\bar{n}_\mathrm{B}$. Comparing the entries in the $S_3$ column of Table \ref{tab:proposed_parameters}, we see that a negative $S_3$ is achievable with modest amounts of additional cooling.

The final two columns of Table \ref{tab:proposed_parameters} give the fraction $\mathcal{F}$ of true positive \{1,0\} clicks for the corresponding experimental parameters. The `resolving' photodetectors column is calculated using Eq. \eqref{eq:false_positives_ratio_resolving} while the `non-resolving' column uses Eq. \eqref{eq:false_positives_ratio_non_resolving}. We have mentioned already that setting the amplitude of the entangling pulse to be $|\alpha|=1$ maximises the probability ${P}_{10}$ of heralding the state $\rho^{\{1,0\}}_{\mathrm{out}}$ (Eq. \eqref{eq:heralding_prob}), therefore the main set of results in these two columns assumes $\alpha=1$. However, in order to be confident that we have successfully heralded state $\rho^{\{1,0\}}_{\mathrm{out}}$, we require $\mathcal{F}$ to be as high as possible. We can experimentally tune $\alpha$ by attenuation, and so the values of $\mathcal{F}$ parentheses have been optimised, having been calculated with the value of $\alpha$ that maximises Eqs. \eqref{eq:false_positives_ratio_resolving} and \eqref{eq:false_positive_prob_non_resolving}, with all other variables in the equations kept constant.

\section{\label{sec:discussion}Discussion}
\subsection{Verification of non-Gaussian entanglement\label{sec:discussion_verification}}

Here we discuss the interplay of the non-Gaussian nature of the state and our ability to detect entanglement. As established in Section~\ref{sec:inseparability_criteria}, the equivalence between $D_5<0$ and Simon's criterion provides a necessary and sufficient condition for the existence of entanglement in Gaussian states. It must also be noted though that $D_5<0$ is only a sufficient criterion for entanglement in non-Gaussian states. As different moments require different time delays between verification pulses they are affected by open-system dynamics to differing degrees. If there is a high degree of environmental interactions across the time-scales we consider this could impede our ability to identify all Gaussian entanglement through use of $D_5$. A sufficiently high quality factor, and low bath occupation $\bar{n}_\mathrm{B}$, ensure that the system is well-isolated from the environment, meaning that a region of state space where $S_3$ is negative but $D_5$ is positive corresponds to states which are guaranteed to be non-Gaussian. On the other hand, if the consequences of an open system do prevent us from identifying Gaussian entanglement with $D_5$, then we can no longer assume merely from studying $D_5$ and $S_3$ that the state is entangled and non-Gaussian. This ambiguity could potentially be resolved using post-processing mitigation schemes that correct for the effects of noise and open-system dynamics on measured moments \cite{Li2017, Temme2017}. Within the parameter set we have explored in Figs. \ref{fig:D5} and \ref{fig:S3}, the region for which $S_3<0$ does not overlap with that where $D_5<0$. As the quality factor considered is large ($Q=10^5$) we can be confident that all the states which fall in the negative (Green) region of Fig. \ref{fig:S3} are non-Gaussian, and this is confirmed by the non-Gaussianity measure shown in Fig. \ref{fig:non_gaussian}.

As discussed in Section~\ref{sec:results}, both the regions of high non-Gaussianity and negative $S_3$ are centred at $\phi=\pi$. However, we are unable to verify entanglement at higher values of $\mu$ using $S_3$ (Fig. \ref{fig:S3}), despite the measure $\delta(\rho^{\{1,0\}}_{\mathrm{out}})$ suggesting that non-Gaussianity increases with $\mu$ (Fig. \ref{fig:non_gaussian}). This is reinforced by Fig. \ref{fig:n_cooling_interpolated} which suggests that there is a cut-off $\mu_\mathrm{c}$ beyond which $S_3$ fails regardless of cooling (even when $\bar{n}=\bar{n}_\mathrm{B}=0$). 

To aid our understanding of the behaviour of $S_3$ in the higher $\mu$ regime, we can examine the analytic formula for $S_3$ calculated for the idealised state $\rho^{\{1,0\}}_{\mathrm{out}}$ with $\bar{n}=0$ (defined in Eq. \eqref{eq:pure_state_1}). Assuming a closed system, $S_3$ is given by:
\begin{equation}
\label{eq:S3_ground_state}
    S_3=-\frac{\mu^6 e^{-\mu^2}}{64(1+e^{-\frac{\mu^2}{2}}\cos \phi)^3}~.
\end{equation}
Notably, this expression is negative for all choices of $\mu$ and $\phi$, approaching 0 from below as $\mu\rightarrow \infty$. Also note that, Eq. \eqref{eq:S3_ground_state} counter-intuitively suggests that at $\phi=\pi$, $S_3$ tends to $-1/8$ as $\mu\rightarrow 0$ which indicates that there is entanglement even when there is no optomechanical coupling. However, this must be considered in conjunction with the heralding probability in Eq. $\eqref{eq:heralding_prob}$ which tends to 0 as $\mu\rightarrow 0$, meaning that such a state cannot be created. 

From studying this ground state behaviour (for a closed system), we see that the magnitude of $S_3$ decays exponentially with increasing $\mu$. Therefore, in the regime where we consider the effects of $\bar{n}$ and and open-system dynamics, at greater values of $\mu$ the possible negative contribution from entanglement detection is too small to compete with these effects causing $S_3$ to become positive. Even at $\bar{n}=\bar{n}_{\mathrm{B}}=0$ there are interactions with the environment present due to finite $\gamma$, thus for coupling strengths greater than $\mu_{\mathrm{c}}$, the scheme is unable to verify entanglement using $S_3$. Indeed, as $Q$ is improved to values greater than $10^5$, the gray region of Fig. \ref{fig:n_cooling_interpolated} shrinks: $\mu_\mathrm{c}$ increases beyond 3.6, and entangled states in environments of even higher $\bar{n}_\mathrm{B}$ can be verified using $S_3$. This analysis suggests our protocol is better suited to parameter regimes where $\mu$ is up to approximately unity, which also corresponds to what is experimentally accessible at present. The experimentally viable systems we have considered \cite{Rossi2018,Wilson2015,Leijssen2017} (see Table \ref{tab:proposed_parameters}) are well within the range of $\mu$ where $S_3$ captures entanglement (Fig. \ref{fig:n_cooling_interpolated}).

In addition to open-system dynamics, additional mechanical modes that couple to the optical field within the bandwidth of the pulsed interaction will lead to unwanted correlations between the mechanical mode of interest and other mechanical modes, and affect the non-Gaussian entanglement criteria used here. In the pulsed optomechanics literature, this unwanted effect was qualitatively noted in Ref.~\cite{Vanner2011} and then such contributions were experimentally investigated for pulsed position measurements~\cite{Vanner2013, Muhonen2019}. This type of multimode contribution was also an experimental factor and carefully studied in Ref.~\cite{Gut2020}. These unwanted contributions may be minimized by engineering the optical and mechanical mode profiles to provide negligible coupling to unwanted mechanical modes, which has, for instance, been used successfully for photonic crystal devices~\cite{LaGala2022}. Additionally, optical trapping based approaches to optomechanics provide excellent coupling to the centre-of-mass motion with few other modes contributing. Moreover, co-levitation~\cite{Rakhubovsky2020, Brandao2020} provides a promising means to implement the `series' optical circuit we have in Fig.~\ref{fig:series}. Lastly, we would like to note that, rather than optimizing the interaction to couple to a single mechanical mode per device, one can utilize this type of pulsed interaction to entangle multiple mechanical modes of a single device. In this direction, bipartite Gaussian entanglement has been explored~\cite{Neveu2021}, and this proposal can also provide tools and techniques for studies of non-Gaussian entanglement in such a single mechanical device regime.

\subsection{Optical effects\label{sec:discussion_optical}}
The following discussion addresses the effect of optical imperfections for the coherent state input case $\ket{\Phi}_{\mathrm{L}}=\ket{\alpha}_1\ket{0}_2$ (as opposed to the single-photon input case which provides more resilience to the effects considered). Optical losses, dark counts, and the entangling pulse amplitude do not directly affect the values of the moments used to calculate $D_5$ and $S_3$, however they do have experimental implications. For example, the entangling pulse amplitude $\alpha$ dictates the probability of heralding the desired mechanical state $\rho^{\{1,0\}}_{\mathrm{out}}$, see Eq. \eqref{eq:heralding_prob}. In the presence of dark counts, the fraction of false positives also depends on $\alpha$, as well as on $\eta$, and $\mathcal{D}$. These false positive events result in mixing of the desired state with the initial state and higher photon-number contributions. This could introduce errors in the measurement of the moments for $D_5$ and $S_3$, which can be readily mitigated by operating in a parameter regime where $\mathcal{F}$ is close to unity ($>\approx$95\%).

The rows (i) to (iv) of Table \ref{tab:proposed_parameters} show that $\mathcal{F}$ is highly dependent on $\mu$. For photon-number-resolving detectors, $\mathcal{F}$ improves with increasing $\mu$ (keeping other experimental parameters constant). From Eq. \eqref{eq:false_positives_ratio_resolving} we see that as $\mu$ decreases, so does the heralding probability ${P}_{10}$ (with $\phi=\pi$ in Eq. \eqref{eq:heralding_prob}), meaning that the desired state $\rho^{\{1,0\}}_{\mathrm{out}}$ is less likely to be created. Once ${P}_{10}$ is small enough that it is comparable to the dark count rate $\mathcal{D}$, dark counts become significant, thus reducing $\mathcal{F}$. By maximising Eq. \eqref{eq:false_positives_ratio_resolving} with respect to $\alpha$, we can compensate for this reduction in $\mathcal{F}$, as demonstrated by the values in parentheses. For the same reasons as the photon-number-resolving case, the reduction in $\mathcal{F}$ for small $\mu$ is also present for non-resolving photodetectors. However, non-resolving photodetectors suffer an additional drawback and thus never outperform resolving photodetectors: as $\mu$ is increased the probability of a multi-photon interaction grows more quickly than one involving a single photon. As a non-resolving detector cannot distinguish between these and a true \{1,0\} click, this reduces $\mathcal{F}$ compared with a resolving detector (this is demonstrated in row (iv)). One can find the optimal $\mathcal{F}$ (shown in brackets in Table \ref{tab:proposed_parameters}) by adjusting $\alpha$ in Eq. \eqref{eq:false_positives_ratio_non_resolving} to balance the competing effects of dark counts and multi-photon events. In a similar manner to $\mu$, increasing $\bar{n}$ both reduces $P_{10}$ (for $\phi=\pi$) and increases the probability of multi-photon interactions. This means that the qualitative effect of $\bar{n}$ on $\mathcal{F}$ can be explained by the same arguments as for $\mu$.

\section{\label{sec:conclusion}Conclusion}
We have proposed a scheme to generate two-mode mechanical Schr{\"o}dinger-cat states using pulsed nonlinear optomechanical interactions in conjunction with a photon-counting heralding scheme. The heralding scheme is based on an optical interferometer set-up, with a variable phase that is used to control the form of the entanglement generated. To verify the presence of entanglement we have introduced an experimental protocol which exploits subsequent pulsed interactions and measurements in order to obtain moments of the bipartite mechanical state. Inseparability criteria can be computed from these moments and we have considered the $D_5$ and $S_3$ criteria. When used together, the two criteria allow us to identify non-Gaussian entangled states. To assess the feasibility of the protocol we have included the key experimental factors including optical losses, detection inefficiency, and dark counts, as well as open-system dynamics.

Our findings indicate that the protocol presented here provides a realistic means of generating non-Gaussian entanglement between two mechanical oscillators. The inseparability criterion $S_3<0$ can verify entanglement in parameter regimes accessible in state-of-the-art experiments with only modest additional cooling required. While the influence of the environment degrades the entanglement, we have demonstrated that for realistic experimental parameters, including a low optomechanical coupling strength, non-Gaussian mechanical entanglement may still be generated and verified.

Furthermore, the experimental verification scheme proposed here enables the measurement of bipartite mechanical moments of arbitrarily high order. While we have focused on extracting quadrature moments used for the two suggested inseparability criteria, any entanglement or non-classicality test that relies on measuring higher-order moments \cite{Walborn2009,ShchukinRichter2005,Zhang2021} can be applied without making any changes to our set-up. 

\ack
We acknowledge useful discussions with Rufus Clarke, M.~S.~Kim, and B.~Lez. This project was supported by the Engineering and Physical Sciences Research Council (EP/T031271/1, EP/L016524/1, and a Doctoral Prize Fellowship awarded to S.Q.), UK Research and Innovation (MR/S032924/1), the Royal Society, the Wallenberg Initiative on Networks and Quantum Information (WINQ), and a Marie Sk{\l}odowska-Curie Action IF programme (101027183). 

\appendix

\section{\label{app:sideband_res}Finite mechanical evolution during entanglement preparation}

The assumption that we operate in the unresolved sideband $\omega_{\mathrm{M}}/\kappa \ll 1$ is a crucial step in the derivation of the optomechanical interaction unitaries both in the generation and verification stages of our protocol. Here we will discuss the interplay between the sideband resolution ratio $\omega_{\mathrm{M}}/\kappa$ and state generation and verification and find that our protocol is robust with respect to this ratio.

The requirement that $\omega_{\mathrm{M}}/\kappa\ll 1$ implies that the cavity fills and empties on a much faster length of time than the mechanical period, and so in the derivation of the state-generation unitary $U=e^{\mathrm{i}\mu a^\dag a X_{\mathrm{M}}}$ it is assumed that $\dot{X}_{\mathrm{M}}=0$. However, if the sideband ratio is relaxed then we have to consider the mechanical evolution during the timescale of the pulse. For the purposes of exploring $U$ we will focus on a single light mode interacting with a single mechanical mode, but these results can be generalised to the multimode unitary $U=\mathrm{e}^{\mathrm{i}\mu a_i^\dag a_i X_{\mathrm{M}_j}}$ which is used in the main text. Starting with the light-mechanics Hamiltonian in a frame rotating at optical drive frequency which is on resonance with the cavity:
\begin{equation}
\label{eq:hamiltonian_res}
    H/\hbar = \omega_{\mathrm{M}}b^\dag b - g_0 a^\dag a (b^\dag +b)~,
\end{equation}
we can formulate the Heisenberg equations of motion for $X_{\mathrm{M}}$ and $P_{\mathrm{M}}$:
\begin{subequations}
\begin{align}
    \dot{X}_{\mathrm{M}}&=\frac{\mathrm{i}}{\hbar}[H, X_{\mathrm{M}}]=\omega_{\mathrm{M}}P_{\mathrm{M}}\\
    \dot{P}_{\mathrm{M}}&=\frac{\mathrm{i}}{\hbar}[H, P_{\mathrm{M}}]=-\omega_{\mathrm{M}}X_{\mathrm{M}}+\sqrt{2}g_0 a^\dag a~.
\end{align}
\end{subequations}
These coupled differential equations can be solved for $X_{\mathrm{M}}(t)$ and $P_{\mathrm{M}}(t)$. Strictly, $a^\dag a(t)$ should be solved using the Heisenberg-Langevin equation that follows from Eq. \eqref{eq:hamiltonian_res} however for purposes of this order-of-magnitude calculation we can set $a^\dag a=1$ which follows from a single photon input. This gives:
\begin{subequations}
\begin{align}
    X_{\mathrm{M}}(t)&=-\frac{\sqrt{2}g_0}{\omega_{\mathrm{M}}}\cos(\omega_{\mathrm{M}}t)+\frac{\sqrt{2}g_0}{\omega_{\mathrm{M}}}\\
    P_{\mathrm{M}}(t)&=\frac{\sqrt{2}g_0}{\omega_{\mathrm{M}}}\sin(\omega_{\mathrm{M}}t)~.
\end{align}
\end{subequations}
Therefore at time $t$, the optomechanical unitary interaction $U$ should displace the mechanical quadratures according to the following transformations: $X_{\mathrm{M}}\rightarrow X_{\mathrm{M}}+\mu'\sin(\omega_{\mathrm{M}}t)$ and $P_{\mathrm{M}}\rightarrow P_{\mathrm{M}}+\mu'\cos(\omega_{\mathrm{M}}t)$,
where $\mu'=2g_0\omega_{\mathrm{M}}^{-1}\sqrt{1-\cos(\omega_{\mathrm{M}}t)}$ is the effective dimensionless coupling strength. Thus, the displacement predicted by the Hamiltonian in Eq. \eqref{eq:hamiltonian_res} is captured by the following unitary:
\begin{equation}
    U=\exp\bigg[\mathrm{i}\mu'\big(\cos(\omega_{\mathrm{M}}t)X_{\mathrm{M}}+\sin(\omega_{\mathrm{M}}t)P_{\mathrm{M}}\big)\bigg]~.
\end{equation}
We can see from this unapproximated unitary that there are two effects which are dependant on the ratio $\omega_{\mathrm{M}}/\kappa$: (i) the effective size of the displacement $\mu'$, such that as $\omega_{\mathrm{M}}/\kappa\ll 1$ is relaxed $\mu'$ is reduced, (ii) the direction of the displacement in phase space. Therefore, without making any assumptions on the size of $\omega_{\mathrm{M}}/\kappa$ we can still generate an entangled bipartite mechanical state (albeit with a different displacement magnitude and direction).

The displacement direction is freely rotating in phase space during the pulsed interaction. This effect can be be accommodated by adjusting the verification times in order perform homodyne measurements relative to the rotated displacement direction. We can keep track of this rotation by introducing the time-dependent variable $X_{\mathrm{M}}^{\theta}=\cos(\omega_{\mathrm{M}}t)X_{\mathrm{M}}+\sin(\omega_{\mathrm{M}}t)P_{\mathrm{M}}$.

\begin{table*}[]%[htbp]
  \caption{Mechanical parameters taken from recent experiments (cf. Table \ref{tab:proposed_parameters}). Parameters considered are optomechanical coupling rate $g_0$, cavity amplitude decay rate $\kappa$, mechanical frequency $\omega_\mathrm{M}$, and sideband resolution ratio $\omega_{\mathrm{M}}/\kappa$. We can see that strong coupling is not required for our protocol. Using the formula $\mu=2\sqrt{2}g_0/\kappa$ we have derived values for $\mu$ in Table \ref{tab:proposed_parameters}. This formula comes from assuming the pulse length $\tau$ satisfies $\omega_{\mathrm{M}}\ll \tau^{-1}\ll \kappa$. In the final column, the percentage reduction in $\mu$ is calculated using the expression Eq. \eqref{eq:error_in_mu} which results from the ratio $\omega_{\mathrm{M}}/\kappa$.}
\makebox[\textwidth][c]{
    %\begin{ruledtabular}
    {
    \begin{tabular}{c|cccc|c}
    \hline\hline
\multicolumn{6}{c}{Mechanical parameters} \\
\midrule
Refs. & $g_0/2\pi$ & $\kappa/2\pi$ & $\omega_\mathrm{M}/2\pi$ & $\omega_\mathrm{M}/\kappa$ & \% reduction in $\mu$\\ \hline 
\midrule
\cite{Rossi2018} & 127~Hz & 15.9~MHz & 1.139~MHz & $7.16 \times 10^{-2}$ & $8.6\times10^{-2}$\\ 
\cite{Wilson2015} & 20~kHz & 0.44~GHz & 4.3~MHz & $9.77\times 10^{-3}$ & $1.6\times 10^{-3}$\\
\cite{Leijssen2017} & 35~MHz & 8.8~GHz & 3.74~MHz & $4.25\times 10^{-4}$ & $3.0\times 10^{-6}$\\
\hline\hline
\end{tabular}%
}
}
\label{tab:extra_parameters}%
\end{table*}%

We now focus on the effect which the sideband parameter has on the magnitude of the displacement $\mu$. Relaxing $\omega_{\mathrm{M}}/\kappa \ll 1$ leads to a reduction in the effective size of $\mu$ which is described by $\mu'$. This is not necessarily a disadvantage as the inseperability criteria outlined in the verification stage are better-suited to smaller values of $\mu$. Therefore, to analyse the effect of the sideband ratio on the size of $\mu'$ we now consider the unitary which acts on the mechanics $U=\exp[\mathrm{i}\mu'X_{\mathrm{M}}^{\theta}]$ given a single-photon has been detected. Let us take $t=C/\kappa$, where $C$ is of order unity and depends on the pulse shape and duration (the details here are unimportant). Inserting the expression for $\mu'$ and Taylor expanding gives:
\begin{equation}
    U=\exp\bigg[\frac{\mathrm{i} \sqrt{2}g_0 C}{\kappa}\bigg(1-\frac{\omega^2_{\mathrm{M}}C^2}{24 \kappa^2}+\ldots\bigg)X_{\mathrm{M}}^{\theta}\bigg]~.
\end{equation}
We can see that for $\omega_{\mathrm{M}}/\kappa\ll 1$ we recover the mechanical displacement after a single photon measurement $U=\mathrm{e}^{\mathrm{i}\mu X_{\mathrm{M}}}$ where $\mu=\sqrt{2} C g_0/\kappa$ (this is used in Section \ref{sec:Entanglement_Protocol}). For a pulse which satisfies $\omega_{\mathrm{M}}\ll\tau^{-1}\ll\kappa$ then $C=2$, and the percentage reduction in $\mu$ to second order is:
\begin{equation}
\label{eq:error_in_mu}
    \text{\% reduction in $\mu$}=\frac{\omega_{\mathrm{M}}^2 }{6 \kappa^2}~.
\end{equation}
This reduction is presented in Table \ref{tab:extra_parameters} for the experimental parameters we consider in Table \ref{tab:proposed_parameters}.

The sideband resolution ratio is also important for the verification stage. The verification stage uses a different wavelength and cavity mode, so will have a different sideband resolution ratio compared with the generation stage. The effect that significant $\omega_\mathrm{M}/\kappa$ has on the linear interaction $U_{\mathrm{V}}=\mathrm{e}^{\mathrm{i}\chi X_{\mathrm{L}}X_{\mathrm{M}}}$ is analyzed in Ref. \cite{Vanner2011}, where it is demonstrated that the correction to $\chi$ is likewise second order in $\omega_{\mathrm{M}}/\kappa$.

\section{Verification set-up}
In this Appendix we detail the key experimental steps required to calculate arbitrary mechanical moments of an entangled state, which are needed to evaluate the inseparability criteria outlined in Section~\ref{sec:inseparability_criteria}. We begin by introducing the notation used to describe the moments, and then we focus on each of the two verification set-ups shown in Figs. \ref{fig:verification_setup} and \ref{fig:series_verification}. For a given mechanical moment we define the order as the sum of the exponents on the operators, e.g. the order of $\braket{X_{\mathrm{M}_1}^pP_{\mathrm{M}_1}^qX_{\mathrm{M}_2}^rP_{\mathrm{M}_2}^s}$ is $d=p+q+r+s$. We use the notation $\mathcal{S} (\braket{X_{\mathrm{M}_1}^p P_{\mathrm{M}_1}^q X_{\mathrm{M}_2}^r P_{\mathrm{M}_2}^s})$ to to denote the sum of all possible permutations of $p$ lots of $X_{\mathrm{M}_1}$-operators, $q$ lots of $P_{\mathrm{M}_1}$-operators etc., taking into account the fact that operators from different oscillators commute. For example, $\mathcal{S} (\braket{X_{\mathrm{M}_1}^2P_{\mathrm{M}_1}X_{\mathrm{M}_2}})=\braket{X_{\mathrm{M}_1}^2P_{\mathrm{M}_1}X_{\mathrm{M}_2}}+\braket{X_{\mathrm{M}_1}P_{\mathrm{M}_1}X_{\mathrm{M}_1}X_{\mathrm{M}_2}}+\braket{P_{\mathrm{M}_1}X_{\mathrm{M}_1}^2X_{\mathrm{M}_{2}}}$. Knowing the value of $\mathcal{S}(\braket{X_{\mathrm{M}_1}^2P_{\mathrm{M}_1}X_{\mathrm{M}_2}})$ and the lower-order moment $\braket{X_{\mathrm{M}_1}X_{\mathrm{M}_2}}$ is sufficient to gain access to $\braket{X_{\mathrm{M}_1}^2P_{\mathrm{M}_1}X_{\mathrm{M}_2}}$, $\braket{X_{\mathrm{M}_1}P_{\mathrm{M}_1}X_{\mathrm{M}_1}X_{\mathrm{M}_2}}$, and $\braket{P_{\mathrm{M}_1}X_{\mathrm{M}_1}^2X_{\mathrm{M}_2}}$ individually as these can be found via the commutation relation $[X_{\mathrm{M}_i},P_{\mathrm{M}_j}]=\mathrm{i}\delta_{ij}$. For example, if our desired moment is $\braket{X_{\mathrm{M}_1}^2P_{\mathrm{M}_1}X_{\mathrm{M}_2}}$, then $\braket{X_{\mathrm{M}_1}^2P_{\mathrm{M}_1}X_{\mathrm{M}_2}}=\mathcal{S}(\braket{X_{\mathrm{M}_1}^2P_{\mathrm{M}_1}X_{\mathrm{M}_2}})/3+\mathrm{i}\braket{X_{\mathrm{M}_1}X_{\mathrm{M}_2}}$. Hence any desired moment -- composed of a sequence of mechanical quadrature operators $\braket{X_{\mathrm{M}_1}^p X_{\mathrm{M}_2}^q X_{\mathrm{M}_2}^r X_{\mathrm{M}_2}^s}$ -- can be written as the sum over the moments of all the distinct permutations of those operators $\mathcal{S}(\braket{X_{\mathrm{M}_1}^p X_{\mathrm{M}_2}^q X_{\mathrm{M}_2}^r X_{\mathrm{M}_2}^s})$, plus lower-order moments (which as we shall see are already known due to the iterative nature of the procedure). This iterative procedure involves finding the moments of all the permutations of sequences of operators belonging to order $d$ (i.e. finding values for all possible sums $\mathcal{S}$ of order $d$, and consequently individual moments) before being able to unlock moments of order $d+1$. The method prescribed can be used to find a moment with arbitrarily large $d$. 

\subsection{\label{app:parallel} Parallel set-up}

Here we will outline the key experimental steps necessary for the entanglement verification (using the set-up in Fig. \ref{fig:verification_setup}) of a mechanical state created by the parallel set-up in Fig. \ref{fig:parallel}. We will first demonstrate that any arbitrary moment of the form $\braket{X_{\mathrm{M}_1}^p P_{\mathrm{M}_1}^q X_{\mathrm{M}_2}^r P_{\mathrm{M}_2}^s}$ can be calculated using our scheme. Then we will show that for some lower-order moments, which appear in the subdeterminants of $D_5$ and $S_3$, a simpler sequence of verification pulses can be used.

\subsubsection{Generalised scheme}\label{app:general} First, let us consider the case where our desired moment is contained within the expression $\mathcal{S}(\braket{X_{\mathrm{M}_1}^pP_{\mathrm{M}_1}^qX_{\mathrm{M}_2}^rP_{\mathrm{M}_2}^s})$. The entangled mechanical state $\rho^{\{1,0\}}_{\mathrm{out}}$ is generated following a \{1,0\} click (see Fig. \ref{fig:parallel}). As discussed in the main text, this mechanical state depends on the phase $\phi$ in the interferometer and so this phase is kept fixed throughout the verification of the state. For generality, we have included $\phi$ in our expressions but it has no relevance in the verification scheme. All other phases $\zeta_l$ (where $l={1,2,3,4}$) are controllable phases. After the state has been created, a verification pulse is then sent into oscillator 1 at time $\tau=0$ followed by a second at $\tau=\pi/(2\omega_\mathrm{M})$. Each pulse interacts sequentially with the mechanical oscillator, and the phase quadratures of the verification pulses transform according to Eq. \eqref{eq:transform2}. These verification pulses have a different wavelength to the entangling pulse to operate in the linearized regime and so spectral filters ensure that after the optomechanical interaction all the verification pulses are diverted away from the photodetectors in Fig. \ref{fig:verification_setup}. The first pulse follows path 1a and is held in a delay line for $\tau=\pi/(2\omega_\mathrm{M})$ while a switch ensures the second, later pulse is directed along 1b and thereby receives an additional phase shift $\zeta_1$. The two pulses are both incident at the first beam splitter simultaneously, which allows us to measure moments comprised of both position and momentum quadratures. Similarly, two verification pulses are also sent into oscillator 2 at time $\tau=0$ and $\tau=\pi/(2\omega_\mathrm{M})$. After the optomechanical interaction with oscillator 2, with the aid of a spectral filter, the first pulse follows path 2a and is held in a delay line while the second, later pulse travels along 2b and experiences a phase shift of $\zeta_2$ at the beam splitter. The light from each oscillator is then combined at two further beam splitters as depicted in Fig. \ref{fig:series_verification}. There is an additional phase at each of these two beam splitters, $\zeta_3$ and $\zeta_4$. After the second set of beam splitters, a homodyne measurement is performed on each of the four output modes, and the phase quadratures are measured. The phase quadratures at the 4 beam splitter ports are given by:
\begin{subequations}
  \begin{align}
      P_{\mathrm{L}_A}&=[\mathrm{e}^{\mathrm{i}\zeta_3}(X_{\mathrm{M}_1}+\mathrm{e}^{\mathrm{i}\zeta_1}P_{\mathrm{M}_1})+\mathrm{e}^{\mathrm{i}\phi}(X_{\mathrm{M}_2}+\mathrm{e}^{\mathrm{i}\zeta_2}P_{\mathrm{M}_2})]\chi/2+P'_{\mathrm{L}_A}~,\label{eq:P_A}\\
      P_{\mathrm{L}_B}&=[\mathrm{e}^{\mathrm{i}\zeta_3}(X_{\mathrm{M}_1}+\mathrm{e}^{\mathrm{i}\zeta_1}P_{\mathrm{M}_1})-\mathrm{e}^{\mathrm{i}\phi}(X_{\mathrm{M}_2}+\mathrm{e}^{\mathrm{i}\zeta_2}P_{\mathrm{M}_2})]\chi/2+P'_{\mathrm{L}_B}~,\label{eq:P_B}\\
      P_{\mathrm{L}_C}&=[X_{\mathrm{M}_1}-\mathrm{e}^{\mathrm{i}\zeta_1}P_{\mathrm{M}_1}-\mathrm{e}^{\mathrm{i}(\phi+\zeta_4)}(X_{\mathrm{M}_2}+\mathrm{e}^{\mathrm{i}\zeta_2}P_{\mathrm{M}_2})]\chi/2+P'_{\mathrm{L}_C}~,\label{eq:P_C}\\
      P_{\mathrm{L}_D}&=[X_{\mathrm{M}_1}-\mathrm{e}^{\mathrm{i}\zeta_1}P_{\mathrm{M}_1}+\mathrm{e}^{\mathrm{i}(\phi+\zeta_4)}(X_{\mathrm{M}_2}-\mathrm{e}^{\mathrm{i}\zeta_2}P_{\mathrm{M}_2})]\chi/2+P'_{\mathrm{L}_D}~.\label{eq:P_D}
  \end{align}  
\end{subequations}
Here, $P'_{\mathrm{L}_k}$ (where $k=\{A,B,C,D\}$) can be understood as the momentum quadrature of the pulse in the absence of the oscillators, and commutes with all mechanical quadrature operators. The momentum quadratures $P'_{\mathrm{L}_k}\propto P_{\mathrm{L}_\mathrm{in}}$; for example, the topmost output in Fig. \ref{fig:verification_setup} has $P'_{\mathrm{L}_A}=(\mathrm{e}^{\mathrm{i}\zeta_3}(1+\mathrm{e}^{\mathrm{i}\zeta_1})+\mathrm{e}^{\mathrm{i}\phi}(1+\mathrm{e}^{\mathrm{i}\zeta_2}))P_{\mathrm{L}_{\mathrm{in}}}/2$. 
In the main text we assumed $P_{\mathrm{L}_\mathrm{in}}$ has vacuum noise statistics, however here we generalise the expressions and assume only that all 4 verification pulses have identical initial statistics, and that their momentum quadratures after each optomechanical interaction are described by Eq. \eqref{eq:transform2}.
The process of creating the entangled states, followed by sending in these four pulses of light and performing a homodyne measurement of the four outputs to measure the momentum quadratures \eqref{eq:P_A}-\eqref{eq:P_D} is repeated over many runs with a fixed set of phases $\{\zeta_1,\zeta_2,\zeta_3,\zeta_4\}$. 
This allows us to build up a data set for each of the homodyne measurements in Eqs. \eqref{eq:P_A}-\eqref{eq:P_D}. We can sum over this data set $\{P_{\mathrm{L}_k}\}_{i=1}^N$ using Eq. \eqref{eq:direct_sum} to extract increasing orders of moments. 
For example, considering Eq. \eqref{eq:P_A} which describes the topmost output in Fig. \ref{fig:verification_setup}, the $d^\mathrm{th}$ order moment of the optical momentum quadrature can be expanded in terms of mechanical quadratures and $P'_{\mathrm{L}_A}$:

\begin{equation}
    \braket{P_{\mathrm{L}_A}^d}=(1/2)^d\sum_{j=0}^d \binom{d}{j} \braket{[\mathrm{e}^{\mathrm{i}\zeta_3}(X_{\mathrm{M}_1}+\mathrm{e}^{\mathrm{i}\zeta_1}P_{\mathrm{M}_1})+\mathrm{e}^{\mathrm{i}\phi}(X_{\mathrm{M}_2}+\mathrm{e}^{\mathrm{i}\zeta_2}P_{\mathrm{M}_2})]^j}(\chi/2)^j\braket{(P'_{\mathrm{L}_A})^{j-d}}~,\label{eq:P_A^d}
\end{equation}
where we have used the fact that $P'_{\mathrm{L}_A}$ commutes with the mechanical quadratures. The value of $\braket{P_{\mathrm{L}_A}^d}$ is calculated summing over the data set of $P_{\mathrm{L}_A}$ measurements according to Eq. \eqref{eq:direct_sum}.
The process also assumes we know $\chi$ accurately and have already obtained full statistics on $P_{\mathrm{L}_{\mathrm{in}}}$ and therefore $P'_{\mathrm{L}_A}$ in an initial calibration stage (we discuss this in Section~\ref{app:calibration}). The remaining terms in the sum of Eq. \eqref{eq:P_A^d} are quadrature moments of order $j=1,\ldots,d$. However, the iterative nature of the process means we have already calculated $\braket{P_{\mathrm{L}_A}^{d-1}}$ and so all mechanical moments up to order $j=d-1$ are known. Therefore, the only unknown moments in the expansion \eqref{eq:P_A^d} are contained in the $d^{\mathrm{th}}$ order term:
\begin{equation}
\begin{split}
&\braket{[\mathrm{e}^{\mathrm{i}\zeta_3}(X_{\mathrm{M}_1}+\mathrm{e}^{\mathrm{i}\zeta_1}P_{\mathrm{M}_1})+\mathrm{e}^{\mathrm{i}\phi}(X_{\mathrm{M}_2}+\mathrm{e}^{\mathrm{i}\zeta_2}P_{\mathrm{M}_2})]^d}=\ldots\\
    &~~~~~~~+\mathrm{e}^{\mathrm{i}p\zeta_3+\mathrm{i}q(\zeta_1+\zeta_3)+\mathrm{i}r\phi +\mathrm{i}s(\zeta_2+\phi)}\mathcal{S}(\braket{X_{\mathrm{M}_1}^p P_{\mathrm{M}_1}^q X_{\mathrm{M}_2}^r P_{\mathrm{M}_2}^s})\\
    &~~~~~~~+\ldots,
    \label{eq:unkown_term}
\end{split}
\end{equation}
where $\mathcal{S}(\braket{X_{\mathrm{M}_1}^p P_{\mathrm{M}_1}^q X_{\mathrm{M}_2}^r P_{\mathrm{M}_2}^s})$ is the term we are trying to calculate. However, the other terms in the sum Eq. \eqref{eq:unkown_term} are of the form $\mathcal{S}(\braket{X_{\mathrm{M}_1}^{\tilde{p}}P_{\mathrm{M}_1}^{\tilde{q}}X_{\mathrm{M}_2}^{\tilde{r}}P_{\mathrm{M}_2}^{\tilde{s}}})$, with $\tilde{p}+\tilde{q}+\tilde{r}+\tilde{s}=d$ and $\{\tilde{p},\tilde{q},\tilde{r},\tilde{s}\}\neq\{p,q,r,s\}$. The coefficients of the terms $\mathcal{S}(\braket{X_{\mathrm{M}_1}^{\tilde{p}}P_{\mathrm{M}_1}^{\tilde{q}}X_{\mathrm{M}_2}^{\tilde{r}}P_{\mathrm{M}_2}^{\tilde{s}}})$ are dictated by the combination of phases chosen $\{\zeta_1, \zeta_2, \zeta_3, \zeta_4\}$. Therefore, we repeat the entire process numerous times with a different set of phases until we obtain the sufficient number of linearly independent equations required to solve for the term  $\mathcal{S}(\braket{X_{\mathrm{M}_1}^p P_{\mathrm{M}_1}^q X_{\mathrm{M}_2}^r P_{\mathrm{M}_2}^s})$. We reiterate that knowledge of the value of $\mathcal{S}(\braket{X_{\mathrm{M}_1}^p P_{\mathrm{M}_1}^q X_{\mathrm{M}_2}^r P_{\mathrm{M}_2}^s})$ is sufficient to calculate the moment of interest via canonical commutation relations. Due to the 4 distinct beam splitter outputs, for a given set of phases $\{\zeta_1, \zeta_2, \zeta_3, \zeta_4\}$, we in fact have access to four linearly independent equations by applying the same analysis to equations \eqref{eq:P_B}-\eqref{eq:P_D}. This reduces the number of distinct sets of phases required by a factor of 4. 

\subsubsection{Special cases}\label{app:special} We will now discuss some special cases of moments that require fewer than 4 verification pulses per run. In fact, all of the moments in the subdeterminants of $D_5$ and $S_3$ fall under this category except $\braket{b_1^\dag b_1 b_2^\dag b_2}$  for which we must use the generalised scheme. 
For single-mode moments $\braket{X_{\mathrm{M}_i}^d(\theta)}$, we require a single verification pulse sent into the $i^\mathrm{th}$ oscillator at time $\tau=\theta/\omega_\textsc{M}$, where $\theta=\{0,\pi/2\}$ (no verification pulses are sent to the $j^{\mathrm{th}}$ oscillator, where $j\neq i$). The entangled state $\rho^{\{1,0\}}_{\mathrm{out}}$ is first created in the usual way (see Fig. \ref{fig:parallel}). For the verification stage, the switches are position 1b and 2b since no delay lines are required. All optical phases $\zeta_l$ are set to 0. Sending in the verification pulse at time $\tau=0$ will allow us to find $\braket{X_{\mathrm{M}_i}^d}$ while sending in the pulse at time $\tau=\pi/(2\omega_\mathrm{M})$ will give us $\braket{P_{\mathrm{M}_i}^d}$. The optical momentum quadratures are measured at each of the four beam splitter outputs, for example $P_{\mathrm{L}_A}(\theta)=\eta_{\mathrm{V}}(P_{\mathrm{L}_{\mathrm{in}}}+\chi X_{\mathrm{M}_i}(\theta))+\sqrt{1-\eta_{\mathrm{V}}}(1+\sqrt{\eta_{\mathrm{V}}})P_V$, where the vacuum noise $P_\mathrm{V}$ has been included from each beam splitter interaction (each beam splitter transmits $\sqrt{\eta_{\mathrm{V}}}$ of the pulse, and each mode passes through two beam splitters). We build up data set of homodyne measurements $\{P_{\mathrm{L}_k}(\theta)\}_{i=1}^N$ and find all moments up to order $d$ using Eq. \eqref{eq:direct_sum}. Rearranging the equation for $\braket{P_{\mathrm{L}_A}^d(\theta)}$ gives:
\begin{equation}
\begin{split}
    \braket{X_{\mathrm{M}_i}^d(\theta)}&=(\eta_{\mathrm{V}} \chi)^{-d}[\braket{P_{\mathrm{L}_A}^d(\theta)}-\sum_{j=0}^{d-1}{d \choose j} (\eta_{\mathrm{V}} \chi)^j\braket{X_{\mathrm{M}_i}^j(\theta)}\braket{(P''_{\mathrm{L}_A})^{d-j}}]~,\label{eq:X^d}
\end{split}
\end{equation}
where $P''_{\mathrm{L}_A}=\eta_{\mathrm{V}} P_{\mathrm{L}_{\mathrm{in}}}+\sqrt{1-\eta_{\mathrm{V}}}(1+\sqrt{\eta_\mathrm{v}})P_{\mathrm{V}}$. Again, this calculation must be done in an iterative way starting with $d=1$ (hence all the moments $\braket{X_{\mathrm{M}_i}^j(\theta)}$, where $j<d$, are presumed to be already known).

For moments of the same oscillator $\braket{X_{\mathrm{M}_i}^pP_{\mathrm{M}_i}^q}$, after creating the entangled state $\rho^{\{1,0\}}_{\mathrm{out}}$ in the usual manner, the switches are moved to position 1a for $i=1$ (or position 2a for $i=2$). Two pulses are sent into the $i^{\mathrm{th}}$ oscillator, one at time $\tau=0$ followed by one at $\tau=\pi/(2\omega_{\mathrm{M}})$ (no pulses are sent to the $j^{\mathrm{th}}$ oscillator). The first pulse is held in a delay line with the switch in position 1a (or 2a). The second, later pulse follows path 1b (or 2b), experiences a $\zeta_{i}$ phase shift, and coincides with the first pulse at the beam splitter. All other phases can be neglected: $\zeta_j=\zeta_3=\zeta_4=0$. For moments like $\braket{X_{\mathrm{M}_1}^p(\theta_1)X_{\mathrm{M}_2}^q(\theta_2)}$ one pulse is sent to oscillator 1 at time $\tau=\theta_1/\omega_{\mathrm{M}}$ and a second pulse is sent to oscillator 2 at $\tau=\theta_2/\omega_{\mathrm{M}}$. The phases $\zeta_1=\zeta_2=0$, but the set $\{\zeta_3,\zeta_4\}$ must be varied. Finally, for a moment such as $\braket{X_{\mathrm{M}_i}^p(\theta_i)X_{\mathrm{M}_j}^q P_{\mathrm{M}_j}^r}$, one pulse is sent into the $i^{\mathrm{th}}$ oscillator at $\tau=\theta_i/\omega_{\mathrm{M}}$ and two pulses are sent into the $j^{\mathrm{th}}$ oscillator at time $\tau=0$ and $\tau=\pi/(2\omega_{\mathrm{M}})$. All four phases in the set $\{\zeta_1, \zeta_2, \zeta_3, \zeta_4\}$ must be varied. With all these special cases, the phase quadratures $P_{\mathrm{L}_k}$ are measured over many runs in order to construct a a data set of homodyne measurement results $\{P_{\mathrm{L}_k}^{(i)}\}_{i=0}^N$ for given set of phases $\{\zeta_1,\zeta_2,\zeta_3,\zeta_4\}$. From the data set all the moments up to $\braket{P_{\mathrm{L}_k}^d}$ are calculated using Eq. \eqref{eq:direct_sum}. The phases within the set are then varied (and we repeat previous step of homodyning over many runs to construct further data sets of homodyne measurements) until we have a sufficient number of linearly independent equations to solve for the term of interest. The mathematics of solving the equations is the same as the general case of $\mathcal{S}(\braket{X_{\mathrm{M}_1}^p P_{\mathrm{M}_1}^q X_{\mathrm{M}_2}^r P_{\mathrm{M}_2}^s})$ and has been examined in detail in \ref{app:parallel}.

\subsubsection{Calibration}\label{app:calibration}
We have assumed throughout \ref{app:general} and \ref{app:special} that we already know the full statistics of $P'_{\mathrm{L}_k}$ where $k=\{A,B,C,D\}$ via a calibration step. This is an important first stage in order to accurately eliminate initial pulse moments from equations such as $\eqref{eq:P_A^d}$. In the absence of any oscillators, we have that $P_{\mathrm{L}_k}=P'_{\mathrm{L}_k}$. Therefore, still using the optical set-up of Fig. \ref{fig:verification_setup}, we send in the verification pulses which we would normally use to obtain a specific moment but in the absence of any optomechanical oscillators (the experimental set-up would need to be adapted further from Fig. \ref{fig:verification_setup} in order to bypass the oscillators). For example, if we are using the generalised regime we would send in 4 verification pulses, 2 in each arm and time $\tau=0$ and $\tau=\pi/(2\omega_\mathrm{M})$. The phases $\zeta_l$ are set to whichever set of phases $\{\zeta_1,\zeta_2,\zeta_3,\zeta_4\}$ we are intending on using for that particular run. Measuring the quadratures $P_{\mathrm{L}_k}$ many times to build up a data set of homodyne measurements from which we can extract $\braket{P'^d_{\mathrm{L}_k}}$ for a given set of $\zeta_l$ phases. 

Another assumption in the verification protocol is precise knowledge of $\chi$, which appears in Eq. \eqref{eq:transform2} and is the interaction strength between the verification pulses and the mechanical system. One can measure $\chi$ precisely in a calibration stage which is separate from the entangling and verification steps. This involves sending pulses of light towards a mechanical state that is in thermal equilibrium and then phase-homodyning the output light. Knowledge of the statistics of the input pulses and the bath temperature allows a value for $\chi$ to be extracted from the value of $\braket{P_{\mathrm{L}}^2}$ which is in turn computed by summing over a data set of many phase-homodyne measurements. 
Furthermore, phase fluctuations will lead to additional phase noise on top of the intrinsic noise of the verification pulses. However, this can seen as reduction in $\chi$ and, provided the variance of this additional phase noise is narrow, this can be accounted in the calibration stage~\cite{Vanner2015}.
\subsection{\label{app:config2}Series set-up}
\begin{figure*}
    \centering
    \includegraphics[width=0.9\textwidth]{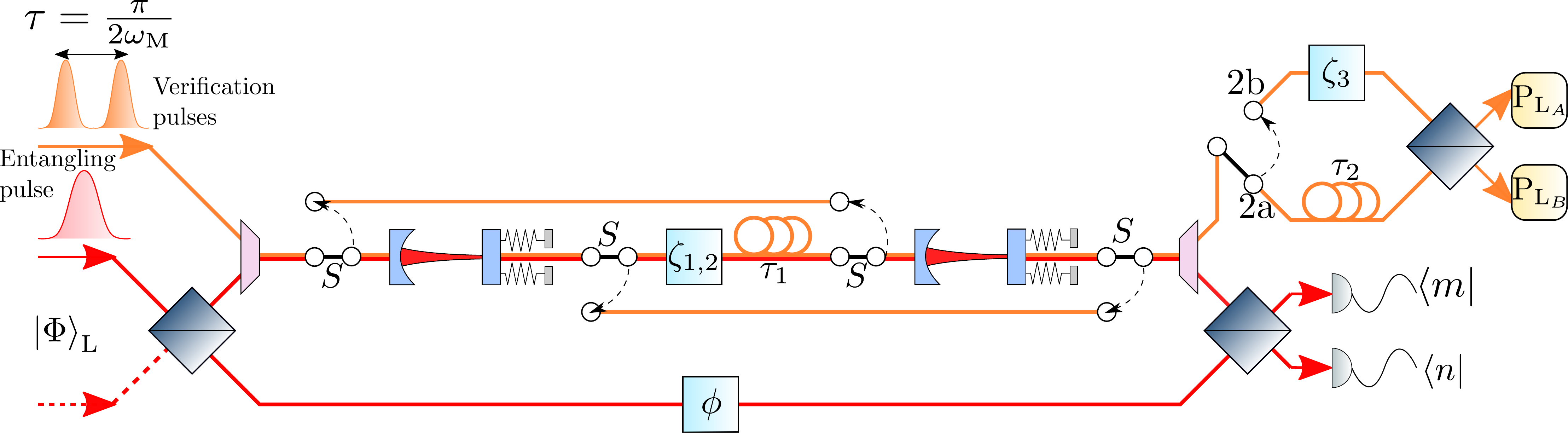}
    \caption{Proposed verification scheme to measure mechanical moments of a state produced via the series set-up. The first stage is to create the entangled state $\rho^{\{1,0\}}_{\mathrm{out}}$ shown by the red path on the diagram; all the switches labelled $S$ are positioned accordingly. After a click event of \{1,0\}, verification pulses shown in orange interact with the mechanical state. Spectral filters ensure that the verification pulses (which have a different wavelength) can follow different paths to that of the entangling pulse. In order to calculate a moment which appears in the sum $\mathcal{S}(\braket{X_{\mathrm{M}_1}^p P_{\mathrm{M}_1}^q X_{\mathrm{M}_2}^r P_{\mathrm{M}_2}^s})$ we send up to two verification pulses per run into the upper arm of the apparatus and then perform homoydne measurements on the verification pulses. This process is repeated many times in order to obtain a data set of homodyne measurements from which moments can be calculated.} The delay line $\tau_1$ is useful if we wish to allow the second oscillator to freely evolve relative to the first oscillator. The switches $S$ allow us to bypass an oscillator in order to get a single-mode mechanical moment. $\tau_2$ allows us to delay the first verification pulse so that the two verification pulses are incident on the final beam splitter coincidentally. The other features of the set-up serve the same purpose as they do in Fig. \ref{fig:verification_setup}.
    \label{fig:series_verification}
\end{figure*}

Our proposed set-up to verify the type of entanglement produced when the two oscillators are in series with each other is displayed in Fig. \ref{fig:series_verification}. In this scheme, we integrate the verification apparatus into the entanglement set-up of Fig. \ref{fig:series}. The mathematical method is identical to the prescribed method for the parallel set-up in \ref{app:parallel}. Sending in up to two verification pulses per run and repeating numerous times allows us to build data sets of homodyne measurements of $P_{\mathrm{L}_A}$ and $P_{\mathrm{L}_B}$. From the data, the moments $\braket{P_{\mathrm{L}_A}^j}$ and $\braket{P_{\mathrm{L}_B}^j}$ for $j=1,\ldots,d$ are directly calculated using Eq. \eqref{eq:direct_sum}. We can expand $\braket{P^d_{\mathrm{L}_A}}$ and $\braket{P^d_{\mathrm{L}_B}}$ in terms of mechanical moments of order $d$ plus lower-order moments. The process is iterative so lower-order moments are already known, leaving only moments of order $d$ to calculate. Repeating with different sets of phases $\{\zeta_1,\zeta_2,\zeta_3\}$ gives us access to a sufficient number of linearly independent equations which can be solved to obtain values such as $\mathcal{S}(\braket{X_{\mathrm{M}_1}^p P_{\mathrm{M}_1}^q X_{\mathrm{M}_1}^r P_{\mathrm{M}_2}^s})$, and consequently any moment that appears in the expansion of $\mathcal{S}(\braket{X_{\mathrm{M}_1}^p P_{\mathrm{M}_1}^q X_{\mathrm{M}_1}^r P_{\mathrm{M}_2}^s})$ can be deduced using canonical commutation relations.

\section{\label{app:decoherence}Langevin equations}
In this Appendix, we outline the method used to model open-system dynamics for the two entangled oscillators. The time-dependent solutions to the quantum Langevin equations \eqref{eq:Langevin_X}-\eqref{eq:Langevin_P} are:
\begin{subequations}
  \begin{align}
  \begin{split}
      X_{\mathrm{M}}(t)&=\mathrm{e}^{-\gamma t/2}[\cos(\omega_{\mathrm{M}} t)+\epsilon\sin(\omega_{\mathrm{M}} t)]X_{\mathrm{M}}(0)+\mathrm{e}^{-\gamma t/2}[\sin(\omega_{\mathrm{M}} t) P_{\mathrm{M}}(0)+\Delta X_{\mathrm{M}}(t)],
      \end{split}
      \label{eq:Langevin_solution_X}\\
      \begin{split}
      P_{\mathrm{M}}(t)&=\mathrm{e}^{-\gamma t/2}[\cos(\omega_{\mathrm{M}} t)-\epsilon \sin(\omega_{\mathrm{M}} t)]P_{\mathrm{M}}(0)+\mathrm{e}^{-\gamma t/2}[-\sin(\omega_{\mathrm{M}} t) X_{\mathrm{M}}(0)+\Delta P_{\mathrm{M}}(t)]~,
      \end{split}
      \label{eq:Langevin_solution_P}
  \end{align}
\end{subequations}
where $\gamma$ and $\omega_{\mathrm{M}}$ are the mechanical damping rate and angular frequency, respectively; and $\epsilon=\gamma/2\omega_{\mathrm{M}}$. We have assumed the optomechanical device has a high quality factor such that $Q=\omega_{\mathrm{M}}/\gamma\gg 1$. The operators $\Delta X_{\mathrm{M}}(t)$ and $\Delta P_{\mathrm{M}}(t)$ contain random excitations entering from the thermal bath:
\begin{subequations}
\begin{align}
    \Delta X_{\mathrm{M}_i}(t)&=\sqrt{2\gamma}\int_0^t dt' \mathrm{e}^{\gamma t'/2}\sin(\omega_{\mathrm{M}}(t-t'))\xi_i(t')~,\label{eq:Delta_X}\\
  \begin{split}
  \Delta P_{\mathrm{M}_i}(t)&=\sqrt{2\gamma}\int_0^t \mathrm{e}^{\gamma t'/2}\xi_i(t')\\&~~~~\times[\cos(\omega_{\mathrm{M}}(t-t'))-\epsilon \sin(\omega_{\mathrm{M}} (t-t'))] \label{eq:Delta_P}
  \end{split}
\end{align}
\end{subequations}
where $\xi_i$ is the Brownian force on the $i^{\mathrm{th}}$ oscillators whose properties are captured in Eqs. \eqref{eq:brownian1} and \eqref{eq:brownian2}.

Since the optomechanical interaction is governed by Eq. $\eqref{eq:transform2}$, the $X_{\mathrm{M}}$ quadrature of a particular oscillator is imprinted on the verification pulse, and so we only make use of Eq. \eqref{eq:Langevin_solution_X}. For example, the moment $\braket{P_{\mathrm{M}_i}}$ is measured at a time that has allowed the $X_{\mathrm{M}_i}$ quadrature to evolve by a quarter of a mechanical oscillation (via the use of a delay line). However, in the presence of damping, the time which is equivalent to a quarter of an mechanical cycle is no longer $\tau=\pi/(2\omega_{\mathrm{M}})$. Instead, it takes time $\tau'$ for the $X_{\mathrm{M}_i}$ quadrature to evolve into the $P_{\mathrm{M}_i}$ quadrature, where $\tau'=\arctan(-\epsilon^{-1})+\pi/\omega_{\mathrm{M}}$. In the limit that $\gamma\rightarrow 0$ then $\tau'\rightarrow \pi/(2\omega_{\mathrm{M}})$ as is expected in the absence of damping. Based on our proposed verification scheme, the value which is measured is in fact $\braket{P_{\mathrm{M}_i}}=\braket{P_{\mathrm{M}_i}(0)}\mathrm{e}^{-\gamma \tau'/2}\sin(\omega_{\mathrm{M}}\sigma \tau')/\sigma $, where $\braket{P_{\mathrm{M}_i}(0)}$ is the expectation of the initial momentum quadrature at the time of state generation. The excitations from the bath via the Brownian force appear in second order terms like $\braket{P_{\mathrm{M}_i}^2}$. These open-system effects have been taken into account for all the matrix elements in $D_5$ and $S_3$ throughout the results presented in Figs. \ref{fig:D5}, \ref{fig:S3}, \ref{fig:n_cooling_interpolated}, and Table \ref{tab:proposed_parameters}.

\section{Optical losses, detector inefficiencies and resolution}\label{app:optical_losses}

\begin{figure}
         \centering
         \includegraphics[height=3.5cm]{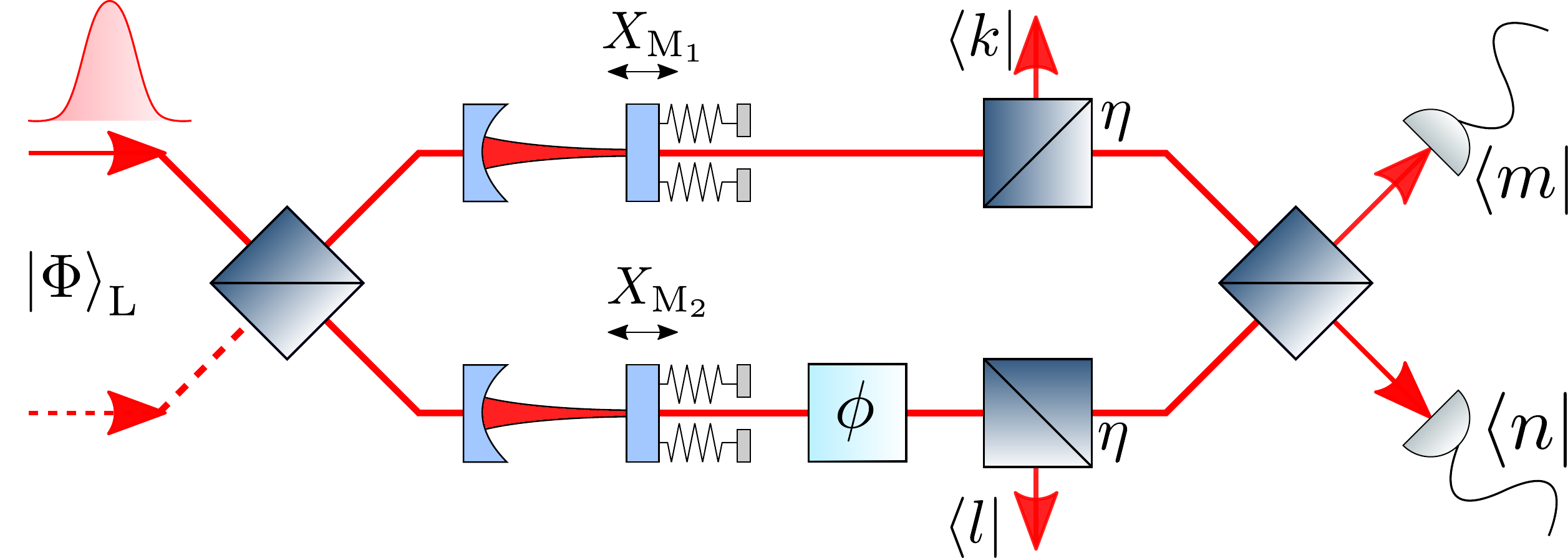}
        \caption{Two beam splitters are introduced into the state generation optical set-up in order to model optical losses and detector inefficiencies (cf. Fig. \ref{fig:parallel}). The beam splitters have intensity transmission $\eta$, and couple the entangling pulse with the environment (which is described by vacuum states). In our formalism, $k$ and $l$ photons are lost to the environment in the upper and lower arms, respectively; $m$ and $n$ photons are measured by the photodetectors. When $\ket{\Phi}_\mathrm{L}$ is a coherent state it is possible that more than one photon interacts with the two oscillators but the photodetectors measure \{1,0\}.}
         \label{fig:parallel_setup_loss}
\end{figure}
In this Appendix we examine how experimental imperfections can affect the success with which we herald our desired mechanical state. We can model optical losses and detector inefficiencies which are present during the state generation stage by introducing loss-model beam splitters with intensity transmission $\eta$ in the lower and upper arms of the interferometer as shown in Fig. \ref{fig:parallel_setup_loss}, such that when $\eta=1$ there are no optical losses and the detectors are perfectly efficient. We introduce $a_3$ and $a_4$ as the annihilation operators of modes of the environment to which entangling pulse couples to at these additional beam splitters. At optical frequencies we can assume that the initial environment mode is well described by the vacuum state. The full measurement operator for the parallel configuration with an injected coherent state is given by:
\begin{equation}
\begin{split}
    \Upsilon_{mnkl}&=\bra{m}_1\bra{n}_2\bra{k}_3\bra{l}_4U_{12}U_{24}U_{13}\mathrm{e}^{\mathrm{i}\mu a^\dag_1a_1X_{\mathrm{M}_1}}\mathrm{e}^{\mathrm{i}\mu a^\dag_2 a_2X_{\mathrm{M}_2}+\mathrm{i} a^\dag_2a_2\phi}U_{12}\ket{\alpha}_1\ket{0}_2\ket{0}_3\ket{0}_4~,
\end{split}
\end{equation}
where $U_{13}$ and $U_{24}$ describe the unitary beam spitter interactions which couple modes 1 and 2 to the environment, with the subscript indicating which modes they act upon:
\begin{subequations}    
    \begin{align}
        U_{13}^\dag a_1 U_{13}=\sqrt{\eta}a_1+\sqrt{1-\eta}a_3,\\
        U_{24}^\dag a_2 U_{24}=\sqrt{\eta}a_2+\sqrt{1-\eta}a_4.
    \end{align}
\end{subequations}
The measurement operator is then:
\begin{equation}
    \begin{split}
        \Upsilon_{mnkl}&=e^{-\frac{|\alpha|^2}{2}}\left(\frac{\sqrt{\eta}\alpha}{2}\right)^{m+n}\left(\frac{\sqrt{1-\eta}\alpha}{\sqrt{2}}\right)^{k+l}\frac{1}{\sqrt{m!n!k!l!}}\\
&~~~~~~\times(e^{\mathrm{i}\mu X_{\mathrm{M}_1}}+e^{\mathrm{i}\mu X_{\mathrm{M}_2}+\mathrm{i}\phi})^m(e^{\mathrm{i}\mu X_{\mathrm{M}_1}}-e^{\mathrm{i}\mu X_{\mathrm{M}_2}+\mathrm{i}\phi})^n\\
&~~~~~~\times e^{\mathrm{i}\mu k X_{\mathrm{M}_1}}e^{\mathrm{i}\mu l X_{\mathrm{M}_2}+\mathrm{i}l\phi},
    \end{split}
\end{equation}
and can be interpreted as losing $k$ and $l$ photons to the environment via each mode (either due to optical losses or detector inefficiencies) but measuring $m$ and $n$ photons at the detectors. 

In addition to the case where $\{k,l\}$ photons are lost to the environment, we have the independent, binomial probability $\mathcal{D}$ that a single dark count is detected during the detection window (we have assumed the probability of more than a single dark count to be negligible on this timescale). With a dark count rate of order 1~s$^{-1}$ and a detection window of order 10~ns, we expect $\mathcal{D}\approx10^{-8}$. 

To calculate the probability of successfully heralding the desired mechanical state given a $\{1,0\}$ click, let ${P}_{mnkl}=\Tr[\Upsilon_{mnkl}\rho_{\mathrm{in}}\Upsilon^\dag_{mnkl}]$, such that ${P}_{mnkl}$ is the probability of losing $\{m,n,k,l\}$ photons at each of the outputs labelled accordingly in Fig. \ref{fig:parallel_setup_loss}. The probability for resolving detectors to herald a mechanical state other than $\rho^{\{1,0\}}_{\mathrm{out}}$, given a \{1,0\} click, is:
\begin{equation}\label{eq:false_positive_prob_resolving}
\begin{split}
    &{P}(\mathrm{False}|\{1,0\} \mathrm{click})=\frac{(1-\mathcal{}{D})\sum_{kl}({P}_{10kl}-{P}_{1000})+\mathcal{D}\sum_{kl}{P}_{00kl}}{(1-\mathcal{}{D})\sum_{kl}{P}_{10kl}+\mathcal{D}\sum_{kl}{P}_{00kl}}~,
\end{split}
\end{equation}
where we have made use of Bayes' Theorem. The probability of heralding the correct $\rho^{\{1,0\}}_{\mathrm{out}}$ state, given a \{1,0\} detector click, is:
\begin{equation}\label{eq:true_positive_prob_resolving}
    {P}(\mathrm{True}|\{1,0\}\mathrm{click})=\frac{(1-\mathcal{D}){P}_{1000}}{(1-\mathcal{}{D})\sum_{kl}{P}_{10kl}+\mathcal{D}\sum_{kl}{P}_{00kl}}~.
\end{equation}
Let us denote the ratio of false positives (Eq. \ref{eq:false_positive_prob_resolving}) to true positives (Eq. \ref{eq:true_positive_prob_resolving}) as $\mathcal{R}$. Then the fraction of true positives we expect for a given number of \{1,0\} clicks is $\mathcal{F}=1/(1+\mathcal{R})$ (shown in Eq. \eqref{eq:false_positives_ratio_resolving}).

We now consider the case where the photodetectors in the heralding stage are non-resolving (a \{1,0\} click would indicate at least 1 photon has been detected in the upper detector, but there could be more photons present). Following the same analysis as above, the analogous equations are:
\begin{equation}\label{eq:false_positive_prob_non_resolving}
\begin{split}
    &{P}(\mathrm{False}|\{1,0\} \mathrm{click})=\frac{\sum_{m>0,kl}{P}_{m0kl}-{P}_{1000}+\mathcal{D}\sum_{kl}{P}_{00kl}}{\sum_{m>0,kl}{P}_{m0kl}+\mathcal{D}\sum_{kl}{P}_{00kl}}~,
\end{split}
\end{equation}
\begin{equation}\label{eq:true_positive_prob_non_resolving}
\begin{split}
    &{P}(\mathrm{True}|\{1,0\} \mathrm{click})=\frac{P_{1000}}{\sum_{m>0,kl}P_{m0kl}+\mathcal{D}\sum_{kl}P_{00kl}}~.
    \end{split}
\end{equation}
The ratio of false positive (Eq. \eqref{eq:false_positive_prob_non_resolving}) to true positive events (Eq. \eqref{eq:true_positive_prob_non_resolving}) is again denoted as $\mathcal{R}$. Using $\mathcal{F}=1/(1+\mathcal{R})$ we can find $\mathcal{F}$ for non-resolving detectors:
\begin{equation}\label{eq:false_positives_ratio_non_resolving}
    \mathcal{F}=\left[\frac{e^{-\eta |\alpha|^2}\mathcal{L}}{\eta {P}_{10}}+\frac{e^{-\eta |\alpha|^2}\mathcal{D}}{\eta {P}_{10}}\right]^{-1}~,
\end{equation}
where ${P}_{10}$ is defined in Eq. \eqref{eq:heralding_prob} (for $\ket{\Phi}_{\mathrm{L}}=\ket{\alpha}_1\ket{0}_2$), and 
\begin{equation}
    \mathcal{L}=\sum_{m=1}^{\infty}\sum_{k=0}^{2m}\binom{2m}{k}\left(\frac{\eta|\alpha|^2}{4}\right)^m\frac{1}{m!}e^{-\mathrm{i}(m-k)\phi}\lambda^{(m-k)^2}~,
\end{equation}
with $\lambda=e^{-\frac{\mu^2}{2}(1+\bar{n}_1+\bar{n}_2)}$.
The results in the `Non-resolving' column in Table \ref{tab:proposed_parameters} are calculated using Eq. \eqref{eq:false_positives_ratio_non_resolving}. When numerically computed, the sum over the $m^\mathrm{th}$ index is truncated at a sufficiently high $m$ such that higher order terms are negligible. Physically, $m$ corresponds to the number of photons at the detector, and since the entangling pulse is a weak coherent state with low mean photon number it is reasonable to neglect very high $m$.

\section{\label{app:experimental_params}Example experimental parameter set}

Here we summarize an example experimental parameter set drawn from \cite{Leijssen2017} together with additional parameters to give the values in the last row of Table \ref{tab:proposed_parameters}.
% and as an example illustrate how all the parameters in the last row of Table \ref{tab:proposed_parameters} are obtained. 
% Table \ref{tab:extra_experimental_parameters} contains experimental parameters taken from Ref. \cite{Leijssen2017} along with further proposed parameters which enable us to calculate $D_5$, $S_3$, and $\mathcal{F}$. 
%
To illustrate the feasibility of our protocol, in Section \ref{sec:results} we focused on the dimensionless, system-independent parameters $\mu$, $Q$, $\bar{n}$, and $\bar{n}_{\mathrm{B}}$. This demonstrated that our protocol of generating and verifying non-Gaussian entanglement is valid for very small $\mu$ of order $10^{-5}$ up to $\mu$ of order $1$.
Even with present-day experiments and only improved cooling we can generate and verify non-Gaussian entangled states.

\begin{table*}[]%[htbp]
  \caption{Example experimental parameter set with key parameters taken from Ref \cite{Leijssen2017}. These parameters demonstrate in detail how $\mu$ and $Q$ are calculated in the last row of Table \ref{tab:proposed_parameters}. Additional experimental parameters are also proposed and listed in order calculate $\bar{n}$ and $\bar{n}_{\mathrm{B}}$, enabling us to calculate $D_5$, $S_3$ and $\mathcal{F}$.}
\makebox[\textwidth][c]{
    %\begin{ruledtabular}
    {
    \begin{tabular}{lll}
    \hline\hline
\multicolumn{3}{c}{Example experimental parameters} \\
\hline \hline
$m$ & Effective mass & 1.5~pg\\
$\omega_{\mathrm{M}}/2\pi$ & Mechanical frequency & 3.74~MHz\\
$x_{\mathrm{zpf}}$ & Zero-point fluctuation $x_{\mathrm{zpf}}=\sqrt{\hbar/2m\omega_{\mathrm{M}}}$ & 43~fm \\
$G/2\pi$ & Parametric coupling $G=\partial \omega_{\mathrm{c}}/\partial x$; cavity resonance frequency $\omega_{\mathrm{c}}$ & 0.8~THz~nm$^{-1}$\\
$g_0/2\pi$ & Optomechanical coupling rate $g_0=G x_{\mathrm{zpf}}$ & 35~MHz\\
$\kappa/2\pi$ & Cavity amplitude decay rate & 8.8~GHz\\
$\gamma/2\pi$ & Mechanical decay rate & 100~Hz\\
$Q$ & Mechanical quality factor $Q=\omega_{\mathrm{M}}/\gamma$ & $3.74\times 10^4$\\
$\mu$ & Dimensionless optomechanical coupling strength $\mu=2\sqrt{2}g_0/\kappa$ & $1.12\times 10^{-2}$\\
\hline \hline
\multicolumn{3}{c}{Additional experimental parameters} \\
\hline \hline
$\bar{n}$ & Initial mechanical phonon occupation number & 0.1 \\
$T_\mathrm{B}$ & Bath temperature & 100~mK\\
$\bar{n}_{\mathrm{B}}$ & Bath occupation number calculated from $T_{\mathrm{B}}$ & 559 \\
$\Delta t$ & Detection window & 10~ns \\
$\lambda_{\mathcal{D}}$ & Dark count rate & 1~s$^{-1}$ \\
$\mathcal{D}$ & Probability of single dark count in $\Delta t$ & $10^{-8}$\\
\hline \hline
\end{tabular}%
}}
\label{tab:extra_experimental_parameters}%
\end{table*}%

\section*{References}
\bibliographystyle{iopart-num}
\bibliography{Bibliography}

\end{document}